\documentclass[]{misc/imag-ms-template}

\title{\go{}: a novel contrast- and resolution-agnostic segmentation tool for Ultra-High Field MRI}

\author{Marc-Antoine Fortin$^{1\ast}$, Anne Louise Kristoffersen$^{1}$, Michael Staff Larsen$^{2}$, \\  Laurent Lamalle$^{3}$, Rüdiger Stirnberg$^{4}$ and Pål Erik Goa$^{1,5}$\\
{\small $^{1}$ Department of Physics, Norwegian University of Science and Technology, Trondheim, Norway}\\
{\small $^{2}$ Department of Computer Science, Norwegian University of Science and Technology, Trondheim, Norway}\\
{\small $^{3}$ GIGA-CRC-Human Imaging, University of Liège, Liège, Belgium}\\
{\small $^{4}$ German Center for Neurodegenerative Diseases (DZNE), Bonn, Germany}\\
{\small $^{5}$ Department of Radiology and Nuclear Medicine, St. Olavs Hospital HF, Trondheim, Norway}\\
{\small $^\ast$Correspondence:  marc.a.fortin@ntnu.no}
}

\usepackage{comment}
\usepackage{epstopdf}
\usepackage{amsthm}
\usepackage{hyperref}
\usepackage{tabularx}
\usepackage{multirow}
\usepackage[table]{xcolor}

\newcommand{\go}{GOUHFI}
\newcommand{\nn}{nnUNet}
\newcommand{\sca}{SCAIFIELD}
\newcommand{\bop}{B$_{1}^{+}$}
\newcommand{\bz}{B$_{0}$}
\newcommand{\tow}{T1w}

\newcommand{\ttw}{T2w}

\newcommand{\pdw}{PDw}

\newcommand{\fsv}{\textit{FastSurferVINN}}

\newcommand{\fss}{\textit{FreeSurfer}}
\newcommand{\sys}{\textit{SynthSeg}}
\newcommand{\ceb}{\textit{CEREBRUM-7T}}

\newcommand{\mpr}{MPRAGE}
\newcommand{\mprr}{MP2RAGE}
\newcommand{\twohi}{(0.25 mm)$^{3}$}
\newcommand{\fori}{(0.4 mm)$^{3}$}
\newcommand{\fivi}{(0.5 mm)$^{3}$}
\newcommand{\sixi}{(0.6 mm)$^{3}$}
\newcommand{\sevi}{(0.7 mm)$^{3}$}
\newcommand{\sevhi}{(0.75 mm)$^{3}$}
\newcommand{\eigi}{(0.8 mm)$^{3}$}
\newcommand{\nini}{(0.9 mm)$^{3}$}
\newcommand{\onei}{1.0 mm$^{3}$}
\renewcommand{\arraystretch}{1.15}

\begin{document}

\maketitle


\begin{abstract}
Recently, Ultra-High Field MRI (UHF-MRI) has become more available and one of the best tools to study the brain for neuroscientists. One common step in quantitative neuroimaging is to segment the brain into several regions, which has been done using software packages like \fss{}, \fsv{} or \sys{}. However, the differences between UHF-MRI and 1.5T or 3T images are such that the automatic segmentation techniques optimized at these field strengths usually produce unsatisfactory segmentation results for UHF images. Thus, it has been particularly challenging to perform region-based quantitative analyses as typically done with 1.5-3T data, considerably limiting the potential of UHF-MRI until now. Ultimately, this underscores the crucial need for developing new automatic segmentation techniques designed to handle UHF images. Hence, we propose a novel Deep Learning (DL)-based segmentation technique called \go{}: Generalized and Optimized segmentation tool for Ultra-High Field Images, designed to segment UHF images of various contrasts and resolutions.\\
For training, we used a total of 206 label maps from four datasets acquired at 3T, 7T and 9.4T. In contrast to most DL strategies, we used a previously proposed domain randomization approach, where synthetic images generated from the 206 label maps were used for training a 3D U-Net. This approach enables the DL model to become contrast-agnostic. \go{} was tested on seven different datasets and compared to existing techniques like \fsv{}, \sys{} and \ceb{}.\\
\go{} was able to the segment six contrasts and seven resolutions tested at 3T, 7T and 9.4T. Average Dice-Sørensen Similarity Coefficient (DSC) scores of 0.90, 0.90 and 0.93 were computed against the ground truth segmentations at 3T, 7T and 9.4T, respectively. These results demonstrated \go{'s} superior performance to competing approaches at each resolution and contrast level tested. Moreover, \go{} demonstrated impressive resistance to the typical inhomogeneities observed at UHF-MRI, making it a new powerful segmentation tool allowing the usual quantitative analysis pipelines performed at lower fields to be applied also at UHF.\\
Ultimately, \go{} is a promising new segmentation tool, being the first of its kind proposing a contrast- and resolution-agnostic alternative for UHF-MRI without requiring fine-tuning or retraining, making it the forthcoming alternative for neuroscientists working with UHF-MRI or even lower field strengths.
\end{abstract}


\section{Introduction}\label{intro}

One of the most important steps in quantitative neuroimaging pipelines is the segmentation of the brain into its different regions. Segmentation can be used to identify specific brain regions for cognitive disease diagnosis, to perform quantitative analyses like relaxometry or volumetry, and to help with surgical planning or image-guided interventions \citep{gonzalez2016review, despotovic2015mri,singh_review_2021, klinger2024reproducibility}. Due to the considerable amount of time and expertise required to produce manual segmentations for many regions and subjects, automatic methods have been developed. Historically, atlas- and Bayesian-based techniques have been proposed, such as MABMIS and \fss{} or FSL-FIRST, respectively \citep{jia2012iterative,fischl2002whole,patenaude2011bayesian}. All propose to automatically segment the whole brain into several cortical and subcortical labels. However, with the developments in graphical processing units (GPU) in the last decade, Deep Learning (DL) has drastically changed the landscape of automatic brain segmentation. Whereas regular machine learning (ML) approaches have shown limited ability to generalize and adapt to complex imaging modalities, convolutional neural networks (CNN) used for DL models have become increasingly successful in handling these challenges \citep{singh_review_2021}. More precisely, the U-Net architecture proposed by \cite{ronneberger2015u} has shown remarkable performance for brain segmentation tasks. Due to its symmetrical encoder-decoder structure with skip connections, creating a U-shaped architecture, the U-Net is able to efficiently extract features at different scales in the images. Recently, several techniques using the U-Net architecture have been proposed like AssemblyNet \citep{coupe_assemblynet_2020}, QuickNat \citep{roy2019quicknat}, SLANT27 \citep{huo20193d}, FastSurferCNN \citep{henschel_fastsurfer_2020} and \fsv{} \citep{henschel_fastsurfervinn_2022}, which allow the segmentation of the brain into more than 25 labels.  

Most of the DL-based brain segmentation techniques, including all of the aforementioned ones, rely on the typical paradigm of using a \tow{} input image with its corresponding segmentation/label map as training data. In order to improve the network capacity to generalize to unseen \tow{} images and increase the training corpus size, extensive data augmentation (DA) is applied on the training data. However, generalization to unseen contrasts and resolutions has shown limitations where segmentation performance quickly decreases when used on images outside the training domain \citep{karani2018lifelong,ghafoorian2017transfer}. This limitation is known as the "domain gap" problem \citep{pan2009survey}. While this issue can be partially addressed by having multi-modality training data or test-time domain adaptation methods \citep{havaei2016hemis,karani2021test}, the network will still struggle when encountering completely unseen images. Alternatively, fine-tuning these models to new contrasts has shown great results. Ultimately, since this fine-tuning is required for every new contrast, this quickly becomes limiting in practice and not an "out-of-the-box" solution. 

Historically, contrast-invariance for brain MRI segmentation has been successfully achieved through Bayesian segmentation \citep{van1999automated}. However, this approach requires considerably more computational time than DL-based techniques. The fastest Bayesian techniques can process one subject in $\sim$15 minutes \citep{puonti2016fast} whereas DL-based techniques require less than 1 minute. Consequently, contrast-invariant Bayesian segmentation techniques have been extremely challenging to implement in clinical settings.

Thus, a novel paradigm for DL training data where randomly generated synthetic images are used instead of real images has emerged. This approach is called domain randomization (DR) and was proposed for brain segmentation for the first time by \cite{billot2023synthseg}. More precisely, synthetic images are generated directly from label maps, using a fully randomized generative model creating images with random contrasts and augmentations that are far beyond what is actually realistic. In \cite{billot2023synthseg}, a novel segmentation tool, \sys{}, was proposed where this DR approach was combined with a 3D U-Net in order to segment MR brain images. \sys{} demonstrated remarkable generalization to unseen contrasts and images with low Signal-to-Noise Ratio (SNR) without the need for fine-tuning or retraining. Moreover, \sys{} outperformed the state-of-the-art Bayesian approach SAMSEG \citep{puonti2016fast} in all tested datasets, in addition to being substantially faster. As a result, the approach proposed by \sys{} has recently been used for other applications like segmentation of white matter (WM) lesions or neonatal brain  \citep{gibson2024segcsvdwmh,valabregue2024comprehensive}, and is widely available through the \fss{} package and distributed with MATLAB (from R2022b and onward).  

While both paradigms (real images + DA versus synthetic images + DR) have been used for many different applications, none of them have been applied to UHF-MRI (i.e., $\geq$7T). UHF-MRI accessibility has increased in the last decade and has even been used for large neuroimaging studies like the Human Connectome Project (HCP) due to its higher SNR, contrast and spatial resolution \citep{trattnig_key_2018}. Despite the several advantages of UHF-MRI, UHF images typically suffer from significant transmit radiofrequency (RF) inhomogeneities compared to lower field strengths, due to the shorter RF wavelength \citep{schick_whole-body_2005}. This results in significant signal and contrast inhomogeneities observed across the image \citep{webb2010parallel}. Although recent developments in parallel transmit (pTx) RF pulses have substantially improved both signal and contrast homogeneity compared to single transmit (1Tx) pulses \citep{gras_universal_2017}, pTx pulses are not widely available and have yet to be applied in large neuroimaging studies. 

This inaccessibility to large datasets with homogeneous UHF images has considerably hindered the development of typical DL-based techniques from \tow{} images. Only one technique, \ceb{}, has been especially designed to segment 7T \tow{} \mprr{} images \citep{svanera_cerebrum7t_2021}. Without retraining or fine-tuning, \ceb{} can segment (0.63mm)$^3$ \tow{} \mprr{} from the Glasgow dataset with a matrix shape of 256$\times$352$\times$224 into six labels: white matter (WM), gray matter (GM), ventricles, basal ganglia, cerebellum and brainstem. Alternatively, considering the limited access to UHF-designed segmentation techniques, several studies have been compelled to use 3T-designed tools like \fss{} on 7T data by implementing extensive preprocessing on the images \citep{zaretskaya2018advantages}. Additionally, \fsv{}, which proposes a solution for sub-millimeter \tow{} images at 3T, has also been recently tested at 7T with pTx \tow{} images and has shown promising results \citep{cabalo2024multimodal,fortin2025}. Ultimately, while tools designed at 3T can be a solution for specific UHF \tow{} images acquired with pTx, they do not provide a reliable solution for most UHF data. Indeed, when both signal and contrast inhomogeneities and resolution differences with 3T are combined, segmentation results are frequently unsatisfactory, requiring important visual quality assurance (QA) and even extremely time-consuming manual corrections.  

Thus, considering the recent increased accessibility of UHF-MRI, there is an urgent need for developing novel automatic segmentation techniques able to address the new issues introduced with UHF-MRI. To the best of our knowledge, no DL technique currently exists to segment (1) \tow{} UHF images in more than 6 labels, (2) highly inhomogeneous 1Tx UHF images, or (3) non-\tow{} contrast UHF images.

In this work, we propose \go{}: Generalized and Optimized segmentation tool for Ultra-High Field Images. By adapting the DR approach proposed in \cite{billot2023synthseg} to the UHF-MRI context and using a state-of-the-art DL architecture with an extensive training corpus, \go{} is able to segment UHF images of various contrasts and resolutions in clinically feasible times without fine-tuning or retraining. More precisely, we present in detail how \go{} was developed and trained, in addition to present its in-depth quantitative and qualitative evaluation against two other segmentation techniques at 3T and 7T. Furthermore, \go{'s} performance against manual delineations at 9.4T and clinical relevance in volumetry measurements between Parkinson's disease patients and healthy controls was evaluated.

\section{Methods}\label{meth}

\subsection{Datasets}

After conducting a comprehensive review of all sub-millimeter MRI datasets freely available online, the eight following datasets were selected for training and testing \go{}. An overview of all these datasets is available in Table \ref{tab:data}.

\begin{table}[h]
    \scriptsize
    \centering
    \caption{Summary of the datasets used for training and/or testing in this work. The table lists the field strength, resolution, contrast, subject type, vendor, usage and number of subjects for each dataset. ASD: Autism Spectrum disorder, PDP: Parkinson's Disease Patients, Tr: Training, Ts: Test.}
    \begin{tabular}{l r r r r r r r}
        \toprule
        Dataset & Field Strength & Resolution & Contrast & Subjects & Vendor & Use & N\\
        \midrule
        HCP-YA & 3T  & \sevi{} & \tow{}/\ttw{} & Healthy & Siemens & Tr & 80/20\\
        SCAIFIELD & 7T (pTx) & \sixi{} & \tow{}, MPM-T1w,-MTw,-PDw & Healthy  & Siemens & Tr/Ts & 31/10\\
        UltraCortex & 9.4T (1Tx)  & \sixi{}/\eigi{} & \tow{} & Healthy  & Siemens & Tr/Ts & 15/12\\
        ABIDE-II ETHZ & 3T & \nini{} & \tow{} & ASD  & Philips & Tr & 34\\
        ABIDE-II EMC & 3T & \nini{} & \tow{} & ASD  & GE & Tr & 46\\
        MPI-CBS & 7T (1Tx) & \fori{} & \tow{} & Healthy  & Siemens & Ts & 28\\
        STRAT-PARK & 7T (1Tx) & \sevhi{} & \tow{} & PDP/Healthy  & Siemens & Ts & 45\\
        CEREBRUM-7T & 7T (1Tx) & (0.63 mm)$^3$ & \tow{} & Healthy & Siemens & Ts & 21\\
        Human Brain Atlas & 7T (1Tx)  & (0.25 mm)$^3$ & \tow{}  & Healthy  & Siemens & Ts & 1\\
        \bottomrule
    \end{tabular}
    \label{tab:data}
\end{table}

\subsubsection{Human Connectome Project: Young Adult}

The Human Connectome Project Young Adult (HCP-YA) \citep{van2012human} is a large neuroimaging study including structural and functional MR images obtained at 3T and 7T on healthy participants between the ages of 22 and 35 years. For \go{}, a subset of 100 randomly selected subjects with preprocessed structural \sevi{} \tow{} \mpr{} and \ttw{} SPACE images acquired at 3T were used. The preprocessing steps included: gradient distortion correction, coregistration and averaging of both \tow{} and \ttw{} runs individually (each sequence is acquired twice per session), Anterior Commissure-Posterior Commissure (ACPC) registration, brain extraction, field map distortion correction, coregistration of \ttw{} to the \tow{} and a bias field correction. More details on the acquisition parameters of both \mpr{} and SPACE sequences and preprocessing steps can be found online\footnote{\url{https://www.humanconnectome.org/storage/app/media/documentation/s1200/HCP_S1200_Release_Reference_Manual.pdf}}. The complete dataset is freely available online\footnote{\url{https://www.humanconnectome.org/study/hcp-young-adult}}. In this work, 80 subjects were used for training and 20 for testing. 

\subsubsection{SpinoCerebellar Ataxias: advanced Imaging with ultra-high-FIELD MRI}

The SpinoCerebellar Ataxias: advanced Imaging with ultra-high-FIELD MRI (\sca{}) is a project aiming at establishing quantitative UHF-MRI biomarkers for polyglutamine SCAs\footnote{\url{https://www.dzne.de/en/research/projects/scaifield/about/}}. For this purpose, a multi-center study has been conducted on 41 healthy participants with data acquired on two 7T MAGNETOM Terra and one 7TPlus MAGNETOM scanners (Siemens Healthineers, Erlangen, Germany) with the same 8Tx/32Rx head coil model in pTx mode. All sequences were acquired using Universal pTx RF Pulses (UP) \citep{gras_universal_2017} created from a database of \bz{} and \bop{} maps acquired at each partner site. The imaging protocol included acquisition of a Multi-Parameter-Mapping (MPM) dataset consisting of Magnetization Transfer-, T1- and Proton Density-weighted multi-echo spoiled gradient echo contrasts and a \tow{} \mpr{}, all at \sixi{} resolution. For this study, 31 subjects were used for training and 10 for testing. The first echo time images of the MPM images were also used for testing (denoted MPM-MTw, MPM-T1w and MPM-PDw).

\subsubsection{UltraCortex}

The UltraCortex \citep{mahler2024ultracortex} is a collaborative project between the Max Planck Institute for biological Cybernetics’ High-Field Magnetic Resonance and University Hospital Tübingen’s Biomedical Magnetic Resonance Departments providing MR images acquired at 9.4T on 78 healthy adult volunteers (M/F: 50/28, age range: 20-53 years old). In total, 86 examinations were performed with either the \mpr{} (n=18) or \mprr{} (n=68) sequence with sub-millimeter resolutions of \sixi{}, \sevi{} and \eigi{}, depending on the subject\footnote{\url{https://www.ultracortex.org/}}. The images were acquired on a 9.4T whole-body MRI scanner (Siemens Healthineers, Erlangen, Germany) with a 16-channel dual-row transmit array operating in CP+ mode paired with a 31-channel receive array. For \mprr{}, the images were \bop{}-corrected and the background noise was removed using the regularization approach proposed in  \cite{o2014robust}. All images were skull-stripped using \textit{SynthStrip} \citep{hoopes2022synthstrip}. Additionally, a set of manual segmentations for WM and GM is provided for 12 subjects which was used as a test dataset for this study (n=8 \sixi{} \mprr{}, n=1 \eigi{} \mprr{} and n=3 \sixi{} \mpr{}). These manual labels were first produced by \fss{}, manually corrected by student assistants and then validated by two expert neuroradiologists. More details on the data acquisition and processing can be accessed in \cite{mahler2024ultracortex}.

\subsubsection{Autism Brain Imaging Data Exchange (ABIDE) II}

The Autism Brain Imaging Data Exchange (ABIDE) II \citep{di2017enhancing} is a large 3T dataset containing 1114 subjects across 19 institutions with different autism spectrum disorders freely available online\footnote{\url{http://fcon_1000.projects.nitrc.org/indi/abide/abide_II.html}}. In this work, two sub-cohorts using \tow{} images at \nini{} resolutions were used. The first one, named ETHZ, included 34 subjects acquired with a 3T Philips Achieva scanner (Philips Healthcare, Best, Netherlands) at ETH Zurich. The second sub-cohort, EMC, acquired 46 subjects with a 3T GE MRI scanner (General Electric Discovery MR750, Milwaukee, MI, USA) at the Erasmus University Medical Center in Rotterdam. More details about the scanning procedure and parameters can be obtained by following the link provided above. All images from both sub-cohorts were used for training.

\subsubsection{Max Planck Institute for Human Cognitive Brain Sciences}

The Open Science CBS Neuroimaging Repository is a dataset repository containing high-resolution and quantitative MRI data acquired at 7T, with single-transmit channel, at the Max Planck Institute for Human Cognitive Brain Sciences (MPI-CBS) in Leipzig \citep{tardif2016open}. The dataset includes 28 \mprr{} images acquired on healthy subjects (M/F: 13/15, age: 26$\pm$4 years old) at \fivi{} but reconstructed at a resolution of \fori{}. All shared images have previously been skull-stripped. 

\subsubsection{STRAT-PARK}

The START-PARK cohort \citep{stige2024strat} is a large ongoing initiative trying to stratify Parkinson's disease (PD) using a multi-disciplinary and multi-center longitudinal cohort composed of PD and neurologically healthy control individuals from Norway and Canada. One branch of STRAT-PARK proposes to use 7T MRI to stratify PD individuals using a high-resolution, multi-contrast and quantitative protocol including both anatomical and functional images. As part of the imaging protocol, a \sevhi{} \mprr{} was acquired with a 7T MAGNETOM Terra scanner (Siemens Healthineers, Erlangen, Germany) using 1Tx channel. The \mprr{} has been skull-stripped in previous work done locally. For this project, a total of 45 subjects were used for testing with 24 PD patients (PDP) (M/F: 13/11, age: 66$\pm$7 years old) and 21 healthy controls (HC) (M/F: 10/11, age: 60$\pm$9 years old).

\subsubsection{\ceb{}: Glasgow dataset}

As part of the work presented in \cite{svanera_cerebrum7t_2021}, the test dataset used to assess \ceb{}'s performance composed of 21 scanning sessions (11 subjects) acquired with a Siemens 7T Terra MAGNETOM scanner at the Queen Elizabeth University Hospital (Glasgow, UK) was made available online\footnote{\url{https://search.kg.ebrains.eu/instances/Dataset/2b24466d-f1cd-4b66-afa8-d70a6755ebea}}. Each session contains a (0.63 mm)$^3$ 1Tx \tow{} \mprr{} and the automatic segmentations computed by \ceb{}. All 21 examinations were used for testing \go{} against \ceb{}.

\subsubsection{Human Brain Atlas}

The Human Brain Atlas (HBA) is an initiative from \cite{schira2023humanbrainatlas} with the goal of creating an \textit{in vivo} atlas of the human brain at (0.25mm)$^3$ resolution from 7T MR images. In order to do so, they have reconstructed a (0.25mm)$^3$ \tow{} \mprr{} from 11 individual \fori{} 1Tx \tow{} \mprr{} scans from the same subject. This single subject, ultra-high resolution reconstructed \mprr{} was used for testing \go{}. More details about the initiative and the data can be found online\footnote{\url{https://osf.io/ckh5t/}}

Each study was approved by the local review boards of each site/institution and participants of the individual studies signed a written informed consent form before scanning. Complete ethic statements are available at each respective study web pages and publications.

\subsection{Data processing}

\subsubsection{Original label map production}

All \tow{} images used in this study were segmented using \fsv{} \citep{henschel_fastsurfervinn_2022} (v2.3.0) with the \textit{--seg\_only} flag in order to produce automatic whole brain segmentations into 35 structures/labels. The list of labels produced by \fsv{} and used in this work, which follows the standard \fss{} lookup table convention \citep{fischl2002whole}, is available in Appendix \ref{appA}.

Since the \tow{} images from the \sca{} and UltraCortex dataset have been acquired at UHF-MRI and were used for training in this study, extensive visual Quality Assurance (QA) has been conducted on all label maps produced by \fsv{}. For \sca{}, the pTx \mpr{} images were N4-corrected \citep{tustison2010n4itk} before being segmented. For both datasets, subjects where low segmentation quality due to motion or important signal inhomogeneities was detected were excluded from the training dataset. For UltraCortex, only 15 of the 78 subjects with \eigi{} \mprr{} images were assessed good enough to be used for training.

\subsubsection{Creation of new label maps and skull-stripped images}

Skull-stripping is a frequent preprocessing step done in most neuroimaging studies. However, non-brain tissues can remain since the skull-stripping quality is highly dependent on many factors like MR scanners, sequences used and image quality \citep{pei2022general,kleesiek2016deep}. In order to simulate a low quality skull-stripping procedure, new label maps with an "extra-cerebral" label were created from the original label maps produced by \fsv{}. First, morphological operations were applied on the brain mask generated by \fsv{}. Binary closing was applied to remove possible holes in the mask and dilation was then applied on the filled mask. The number of voxels used for dilation was 4 (UltraCortex and ABIDE-II) or 5 (HCP and SCAIFIELD) depending on the resolution of the label maps. These two steps were executed to correct in case of too stringent brain masking. This new mask was then used together with the original mask to create the new extra-cerebral label by assigning all voxels mutually exclusive to the new mask as this new extra-cerebral label (i.e., voxels present in the new mask but not in the original one). Then, the new label map was created by assigning this new label value to the corresponding voxel position in its original label map.

Once the new mask was created, the corresponding input \tow{} image from which the label map was created was masked to create a skull-stripped version. For multi-contrast datasets like the HCP and SCAIFIELD, all contrasts were coregistered before applying the masking. The rationale of using skull-stripped training data was based on (1) the fact that some of the datasets were directly shared as skull-stripped data, (2) the inaccessibility to the whole-head segmentation algorithm used in \sys{} and (3) the lack of signal outside the brain for UHF-MRI data, making intensity-based whole-head segmentation substantially harder than at 3T.
 
\subsection{Generation of synthetic training images}

As described in section \ref{intro}, the core concept behind \sys{} is the creation of a training dataset composed solely of synthetic images randomly generated from label maps, regardless of whether they were generated automatically or manually \citep{billot2023synthseg}. In order to create synthetic images, the generative model uses fully randomized parameters from predefined priors, generating images with random (and unrealistic) contrasts, morphologies, artifacts, and noise levels. For an exhaustive explanation behind the generative model developed in \sys{}, we recommend the reader to consult \cite{billot2023synthseg}, since the focus of this section is towards variations from the original model. 

In this work, the generative model used in \sys{} was adapted for UHF images. More precisely, the parameter simulating bias field in the synthetic images was increased from 0.6 to 0.9 to simulate the large signal inhomogeneities frequently observed at UHF, and all parameters related to the randomized downsampling of the synthetic images were disabled to allow the creation of synthetic images at native sub-millimeter resolution. Contrary to the original generative model used for \sys{} where they were excluded, the choroid plexus (both hemispheres), cerebrospinal fluid (CSF) and WM-hypointensities labels were included to generate synthetic images. The extra-cerebral label was also included in the generative model to synthesize the images, but was excluded from the final label maps. The rest of the generative model was kept as originally proposed and one synthetic image was generated per training label map. An example case is shown in Figure \ref{fig:pipe}, and the parameters of the generative model used in this study are provided in Appendix \ref{appB}.

\begin{figure}[htbp]
    \begin{center}
        \includegraphics[width=0.95\textwidth]{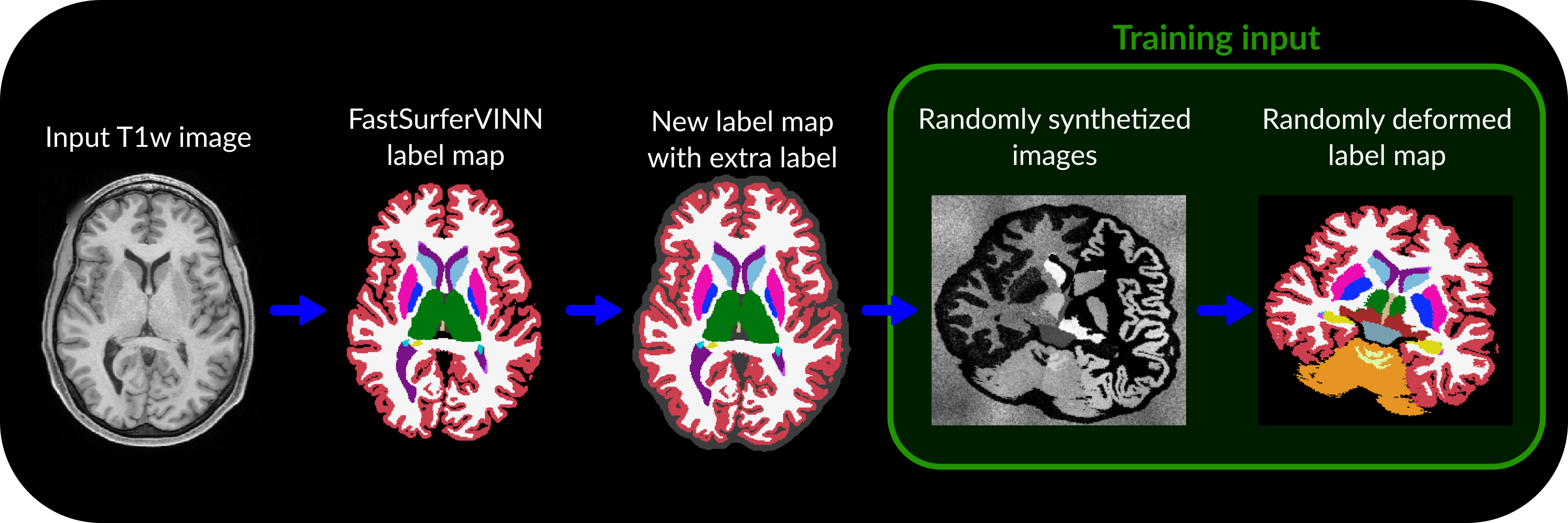}
        \caption{Pipeline used to create the training data for \go{}. To produce the training data, the sub-millimeter \tow{} image was used as input to \fsv{} in order to create a label map of the whole-brain with 35 labels. A new label map was then created by modifying the \fsv{} output by adding an extra-cerebral label based on the morphologically modified brain mask (dark gray area surrounding the cerebral cortex on the third sub-figure). This new label map was then used as input to the modified generative model from \sys{} to create the randomly deformed version of it. Then, the augmented label map was used to generate the synthetic image where, as explained in \cite{billot2023synthseg}, a mean and standard deviation are randomly sampled from a normal distribution to generate a noisy signal for each label iteratively. Ultimately, the extra-cerebral label was kept for generating the synthetic images but excluded from the final label map in order to simulate signal surrounding the cortex like CSF or remaining non-brain tissues from low quality skull-stripping. Finally, the generated synthetic image with its corresponding deformed label map (from which it was generated) were used as the training data (green box).}
        \label{fig:pipe}
    \end{center}
\end{figure}

\subsection{Deep Learning model}

\subsubsection{Training data}

The training dataset used was composed of 206 different subjects from the HCP (n=80), SCAIFIELD (n=31), UltraCortex (n=15) and ABIDE-II (n=80) datasets with 80\% of the subjects randomly assigned as training with the remaining 20\% as validation using 5-fold cross-validation. One aspect to achieve contrast-agnosticity is the use of real MR images of different contrasts for the validation set, while the synthetic images are used for training only. In other words, if one subject was assigned to the validation set, the input \tow{} image and corresponding label map were used and not the synthetic image produced by the generative model. Consequently, for the UltraCortex and ABIDE-II datasets, the \tow{} images with their corresponding label maps were used for validation, whereas for the HCP and SCAIFIELD, the \ttw{} and \pdw{} images were used respectively (both coregistered to their corresponding \tow{} image used for the label map creation).     

\subsubsection{Network architecture}\label{meth:network}

In this work, the \nn{} framework (v2.4.1) with residual encoder \citep{isensee2021nnu,isensee2024nnu} was used to implement the DL model. The "3D U-Net full image resolution" with "Large Presets" was selected as the configuration, considering the isotropic and sub-millimeter nature of the training data and hardware limitations. More precisely, a 3D U-Net composed of six layers with 32, 64, 128, 256, 320 and 320 features, using Leaky ReLU \citep{maas2013rectifier}, 3$\times{}$3$\times{}$3 kernel size for convolutions, with a patch size of 192$\times$192$\times$160 and a batch size of 2 were used. The loss function was the sum of the Dice and cross-entropy losses \citep{drozdzal2016importance}. 

The data augmentation step performed by default by the \nn{} was disabled, since extensive data augmentation was already applied by the DR approach used in the generation step of the synthetic images. Since the training dataset was composed of several resolutions, the median resolution of the training dataset (\sevi{}) was used as the training resolution for the network. Consequently, images and label maps at different resolutions were externally up- or down-sampled to the median resolution before being fed to the network. For images, 3D cubic spline interpolation was used, whereas 3D linear interpolation with one-hot encoding was used for label maps.  

The complete description of all steps performed by the \nn{} framework can be found in \cite{isensee2021nnu}. 

\subsubsection{Training setup}

The model was trained for a total of 500 epochs for each fold, where one epoch was defined as 250 random mini-batches fed to the network. The AdamW optimizer \citep{loshchilov2017decoupled} was utilized with a base learning rate (LR) of 3$\times$10$^{-4}$, decaying following the poly LR scheduler. The total training time was 6 days (1.2 days/fold) using an NVIDIA Ampere A40 GPU with 48 Gb of VRAM.

\subsubsection{Inference}

Since 5-fold cross-validation was used (i.e., 5 separate models were trained), an ensembling strategy, where the softmax outputs (i.e., probability for each label) from all models are averaged together, was used to produce a single output label map at inference. Moreover, after training all five models, the default post-processing step proposed by \nn{} of keeping only the largest component for each label was tested with the validation dataset. For all labels, the average Dice score was computed with only the largest component compared to the value with all components for that label. Then, if the average Dice score was improved, the post-processing step would be executed for this label on any data to be inferred. For \go{}, this step was applied to all 35 labels except the Left- and Right-Inferior-Lateral-Ventricles.

Since \go{} was trained with segmentations produced from \fsv{}, and U-Nets are sensitive to spatial localization, all data needed to be reoriented to the Left-Inferior-Anterior (LIA) orientation before being inferred. Moreover, the data to be inferred needed to be resampled to the training resolution of \sevi{} before being processed by the network. The data was then resampled back to native resolution after being segmented. Inference, up/downsampling (if using a different resolution from the one used for training) and post-processing took approximately 60 seconds per 3D volume/image. All these steps are implemented in \go{} and all results shown in this work were computed with \go{} version 1.1.0 available at \url{https://github.com/mafortin/GOUHFI/releases/tag/1.1.0}.  

\subsection{Evaluation metrics}

\subsubsection{Quantitative evaluation}

In order to assess the quality of the segmentations produced by \go{}, the Dice-Sørensen Similarity Coefficient (DSC) \citep{dice1945measures,sorensen1948method}, which measures the overlap between two segments (with a value of 1 being a perfect overlap between the two segments) was computed with the following equation: 

\begin{equation}
    DSC = \frac{2 \times |G \cup P|}{G + P}
\end{equation}

where G is the ground truth segment, and P is the predicted segment to be compared.

Moreover, the Average Surface Distance (ASD) \citep{reinke2024understanding}, where a value of 0 represents a perfect alignment of both surfaces evaluated, was computed with the following equation:  

\begin{equation}
    ASD = \frac{\displaystyle\sum_{i=1}^{N_G} d_{G\rightarrow P,i} + \sum_{i=1}^{N_P} d_{P\rightarrow G,i}}{N_G + N_P}\quad .
\end{equation}

Herein, $d_{G\rightarrow P,i}$ is the distance from point $i$ on the surface of the ground truth segment to its nearest point on the surface of the predicted segment; $d_{P\rightarrow G,i}$ is the distance from point $i$ on the surface of the predicted segment to its nearest point on the surface of the ground truth segment; $N_G$ and $N_P$ are the total number of points on the ground truth and predicted surfaces respectively.

Except for the UltraCortex dataset where manual segments were available and the Glasgow dataset where the "inaccurate ground truth" (iGT) was used, the ground truth segmentations in this work were obtained by running \fsv{} on the sub-millimeter \tow{} images for the HCP-YA and the SCA-\tow{} test datasets. For SCA-MPM contrasts, \sys{} was used as the ground truth. Thus, in these specific cases, the ground truth was considered a "silver standard" and a perfect DSC of 1.0 was not necessarily desired, especially in cases where \fsv{} was prone to face difficulties (e.g., $>$3T or $<$\sevi{} resolution). Nevertheless, \fsv{} has previously shown robustness to resolutions $<$\sevi{} and pTx UHF-MRI data \citep{fortin2025}. Both \fsv{} and \sys{} are also the only "out-of-the-box" solutions to quantitatively evaluate \go{} against in most cases since \ceb{} requires retraining outside its training domain for every test dataset, and that producing manual delineations for 35 labels for many subjects was outside the scope of this work. For \ceb{}, the iGT was created by a combination of \fss{}, \textit{AFNI3dSeg} \citep{cox1996afni} and methods from \cite{fracasso2016lines} as described in \cite{svanera_cerebrum7t_2021}. 

In addition, while \fsv{} produces segmentations at the native input image resolution, \sys{} solely segments images at \onei{}, irrespective of the input resolution of the images. In order to obtain label maps at the same resolution as \go{} and \fsv{} and allow for quantitative comparisons, the same external up-sampling strategy using one-hot encoded 3D linear interpolation for label maps as done for \go{} was implemented in-house for \sys{}. 

For calculating DSC and ASD, the choroid plexus (both hemispheres) and WM-hypointensities were excluded since \sys{} does not segment these labels. Consequently, the lateral and inferior lateral ventricles (both hemispheres) were also excluded since both regions are directly impacted by the presence of the choroid plexus label. Finally, the CSF was also excluded since \sys{} defines CSF in a completely different way than \fsv{} and \go{}. In case a label was missing in a label map (i.e., not segmented), DSC and ASD values were set to 0 and NaN respectively. 

\subsubsection{STRAT-PARK: Volumetry analysis}

To further assess \go{}'s performance, a volumetric analysis was performed with the START-PARK dataset. For both HC and PDP, the median group volume for the putamen, amygdala and hippocampus, normalized by the total intracranial volume (TIV), were computed based on the segmentations produced by \fsv{}, \go{} and \sys{}. The TIV values were computed with SPM12 \citep{ashburner2012spm} and the region-of-interest (ROI) volumes were computed by multiplying the voxel volume by the number of voxels for each ROI. These ROIs were selected based on the literature for PD \citep{junque2005amygdalar,pieperhoff2022regional,geng2006magnetic}. A Mann-Whitney U test \citep{mann1947test} with Bonferroni-corrected p-values \citep{bonferroni1936teoria} was computed to measure the statistical differences between each group for both segmentation tools.

\section{Results}\label{res}

\subsection{HCP-YA: Benchmarking against \fsv{} \& \sys{} at 3T}\label{res:hcp}

\begin{figure}[htbp]
    \begin{center}
        \includegraphics[width=0.75\textwidth]{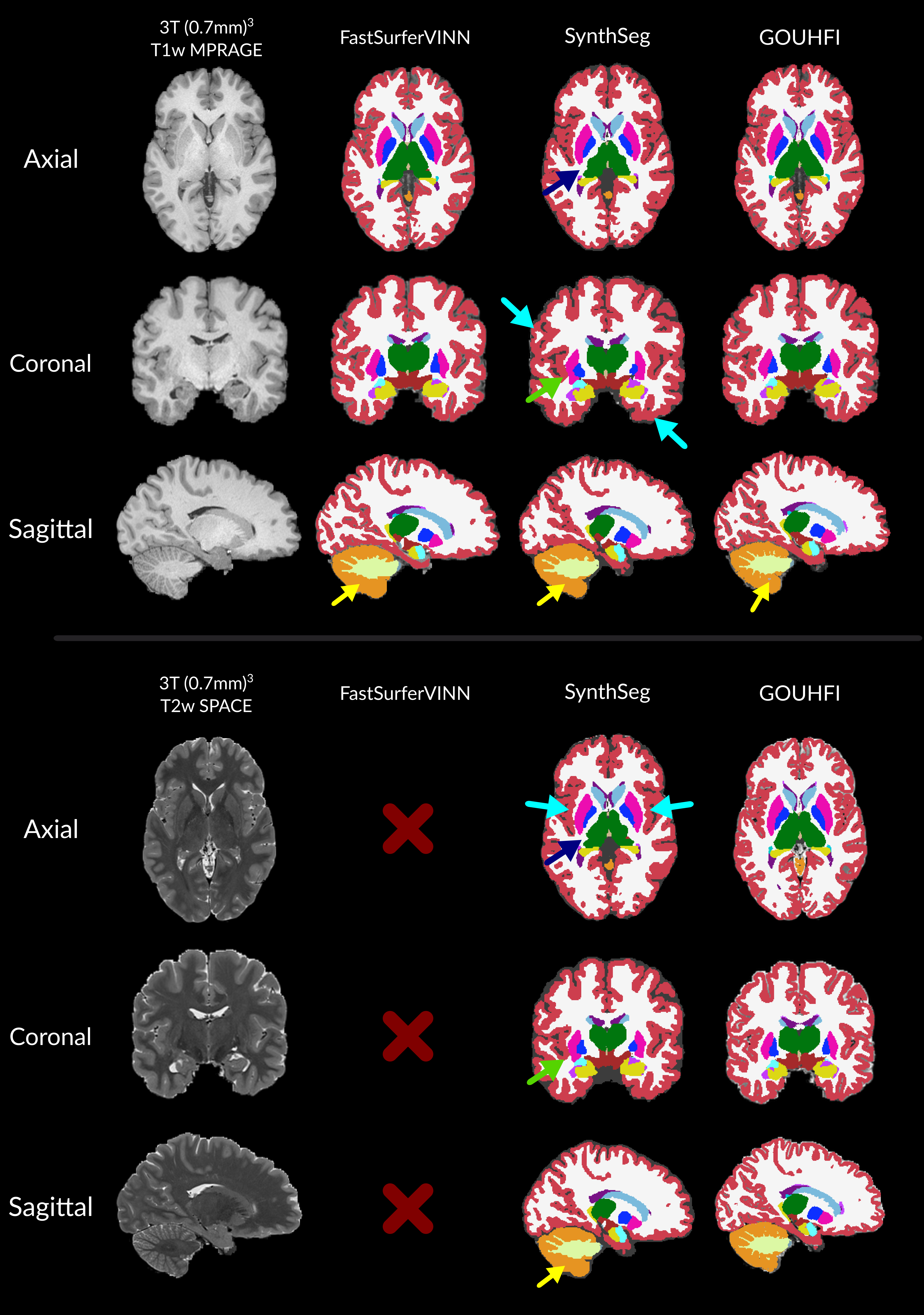}
        \caption{Segmentations produced by \fsv{} (second column), \sys{} (third column) and \go{} (right column) for one subject in all anatomical planes for the \tow{} (top) and \ttw{} (bottom) contrasts from the HCP dataset (3T). No segmentations are shown for the \ttw{} image for \fsv{} since it only segments \tow{} images. Dark blue arrows represent regions of discrepancies for the thalamus region with the ground truth for \sys{}. Turquoise arrows point to cortical regions with limited delineation by \sys{}. Green arrows show systematic errors with \sys{} including the claustrum in the putamen label. Yellow arrows show differences in cerebellum WM segmentation. The labels shown and their colors correspond to the \fss{} lookup table.}
        \label{fig:hcp-segs}
    \end{center}
\end{figure}

The segmentations produced by \sys{} and \go{} for both \tow{} and \ttw{} \sevi{} 3T images and by \fsv{} for the \tow{} images only are shown in Figure \ref{fig:hcp-segs}. No segmentation is shown for the \ttw{} images for \fsv{} since the technique segments \tow{} images only. Visually, segmentations produced on both the \tow{} and \ttw{} images were extremely similar to the ones produced by \fsv{} for both \sys{} and \go{}. Visually, differences in cortex and cerebellum WM delineations could be observed for \sys{} compared to \go{}. Moreover, the segmentation boundary of the putamen was extended laterally for \sys{} compared to both \fsv{} and \go{}. For \sys{}, the thalamus delineation was different following a more irregular boundary than the two other techniques as seen with the axial plane. The median and 95\% confidence intervals (CI) for the DSC and ASD computed across all subjects (n=20) and labels with a selected subset are shown in Table \ref{tab:dsc-asd-hcp}. Overall, \go{} produced higher and lower median DSC and ASD values respectively for all labels except for the cerebellum WM and cortex compared to \sys{}.

\begin{table}[h]
\centering
\begin{threeparttable}
\caption{Median DSC and ASD values (with 95\% CIs) computed for HCP subjects (n=20) using \go{} and \sys{}. The ground truth is the segmentation produced by \fsv{} at native resolution using the \tow{} images. The highest DSC (and lowest ASD) value is shown in bold.}
\scriptsize
\renewcommand{\arraystretch}{1.3} 

\newlength{\mycolwidth}
\setlength{\mycolwidth}{2.4cm}

\begin{tabular}{l
  >{\centering\arraybackslash}p{\mycolwidth}
  >{\centering\arraybackslash}p{\mycolwidth}
  >{\centering\arraybackslash}p{\mycolwidth}
  >{\centering\arraybackslash}p{\mycolwidth}}
\toprule
\textbf{DSC} & \multicolumn{2}{c}{HCP-T1w} & \multicolumn{2}{c}{HCP-T2w} \\
            & GOUHFI & SynthSeg & GOUHFI & SynthSeg \\
\midrule
WM        & \textbf{0.96} [0.96, 0.97] & 0.93 [0.92, 0.93] & \textbf{0.95} [0.94, 0.95] & 0.90 [0.90, 0.91] \\
Cortex    & \textbf{0.92} [0.92, 0.92] & 0.88 [0.88, 0.88] & \textbf{0.90} [0.90, 0.90] & 0.85 [0.85, 0.85] \\
Putamen   & \textbf{0.94} [0.93, 0.94] & 0.90 [0.89, 0.90] & \textbf{0.92} [0.91, 0.92] & 0.88 [0.88, 0.89] \\
Thalamus  & \textbf{0.93} [0.92, 0.93] & 0.91 [0.91, 0.92] & \textbf{0.92} [0.91, 0.92] & \textbf{0.92} [0.92, 0.92] \\
Pallidum  & \textbf{0.85} [0.84, 0.86] & 0.82 [0.81, 0.83] & \textbf{0.84} [0.83, 0.85] & 0.81 [0.79, 0.81] \\
Cerebellum WM & 0.87 [0.87, 0.87] & \textbf{0.88} [0.88, 0.89] & 0.85 [0.85, 0.85] & \textbf{0.87} [0.87, 0.87] \\
Cerebellum Cortex & 0.90 [0.89, 0.90] & \textbf{0.93} [0.92, 0.93] & 0.89 [0.88, 0.89] & \textbf{0.91} [0.90, 0.91] \\
\rowcolor{gray!10}
\textbf{Median (27 labels)} & \textbf{0.91} [0.90, 0.91] & 0.89 [0.88, 0.89] & \textbf{0.89} [0.88, 0.89] & 0.87 [0.86, 0.87] \\
\bottomrule
\end{tabular}

\vspace{1.5em}

\begin{tabular}{l
  >{\centering\arraybackslash}p{\mycolwidth}
  >{\centering\arraybackslash}p{\mycolwidth}
  >{\centering\arraybackslash}p{\mycolwidth}
  >{\centering\arraybackslash}p{\mycolwidth}}
\toprule
\textbf{ASD [mm]} & \multicolumn{2}{c}{HCP-T1w} & \multicolumn{2}{c}{HCP-T2w} \\
                  & GOUHFI & SynthSeg & GOUHFI & SynthSeg \\
\midrule
WM        & \textbf{0.19} [0.19, 0.19] & 0.34 [0.33, 0.34] & \textbf{0.27} [0.27, 0.29] & 0.41 [0.40, 0.42] \\
Cortex    & \textbf{0.26} [0.26, 0.33] & 0.36 [0.36, 0.37] & \textbf{0.31} [0.31, 0.33] & 0.42 [0.42, 0.43] \\
Putamen   & \textbf{0.32} [0.32, 0.34] & 0.54 [0.53, 0.57] & \textbf{0.40} [0.40, 0.43] & 0.62 [0.60, 0.66] \\
Thalamus  & \textbf{0.52} [0.51, 0.57] & 0.56 [0.55, 0.60] & 0.61 [0.59, 0.65] & \textbf{0.54} [0.54, 0.58] \\
Pallidum  & \textbf{0.54} [0.51, 0.58] & 0.56 [0.55, 0.61] & \textbf{0.58} [0.57, 0.62] & 0.62 [0.62, 0.69] \\
Cerebellum WM & 0.67 [0.63, 0.70] & \textbf{0.54} [0.51, 0.59] & 0.81 [0.79, 0.86] & \textbf{0.62} [0.60, 0.67] \\
Cerebellum Cortex & 0.90 [0.87, 0.95] & \textbf{0.57} [0.56, 0.59] & 0.93 [0.91, 0.99] & \textbf{0.66} [0.64, 0.68] \\
\rowcolor{gray!10}
\textbf{Median (27 labels)} & \textbf{0.45} [0.43, 0.47] & 0.50 [0.48, 0.50] & \textbf{0.47} [0.51, 0.54] & 0.55 [0.54, 0.56] \\
\bottomrule
\end{tabular}

\label{tab:dsc-asd-hcp}
\end{threeparttable}
\end{table}

\subsection{SCAIFIELD: Contrast- and resolution-agnostic performance at 7T}\label{res:sca}

\begin{figure}[htbp]
    \begin{center}
        \includegraphics[width=0.65\textwidth]{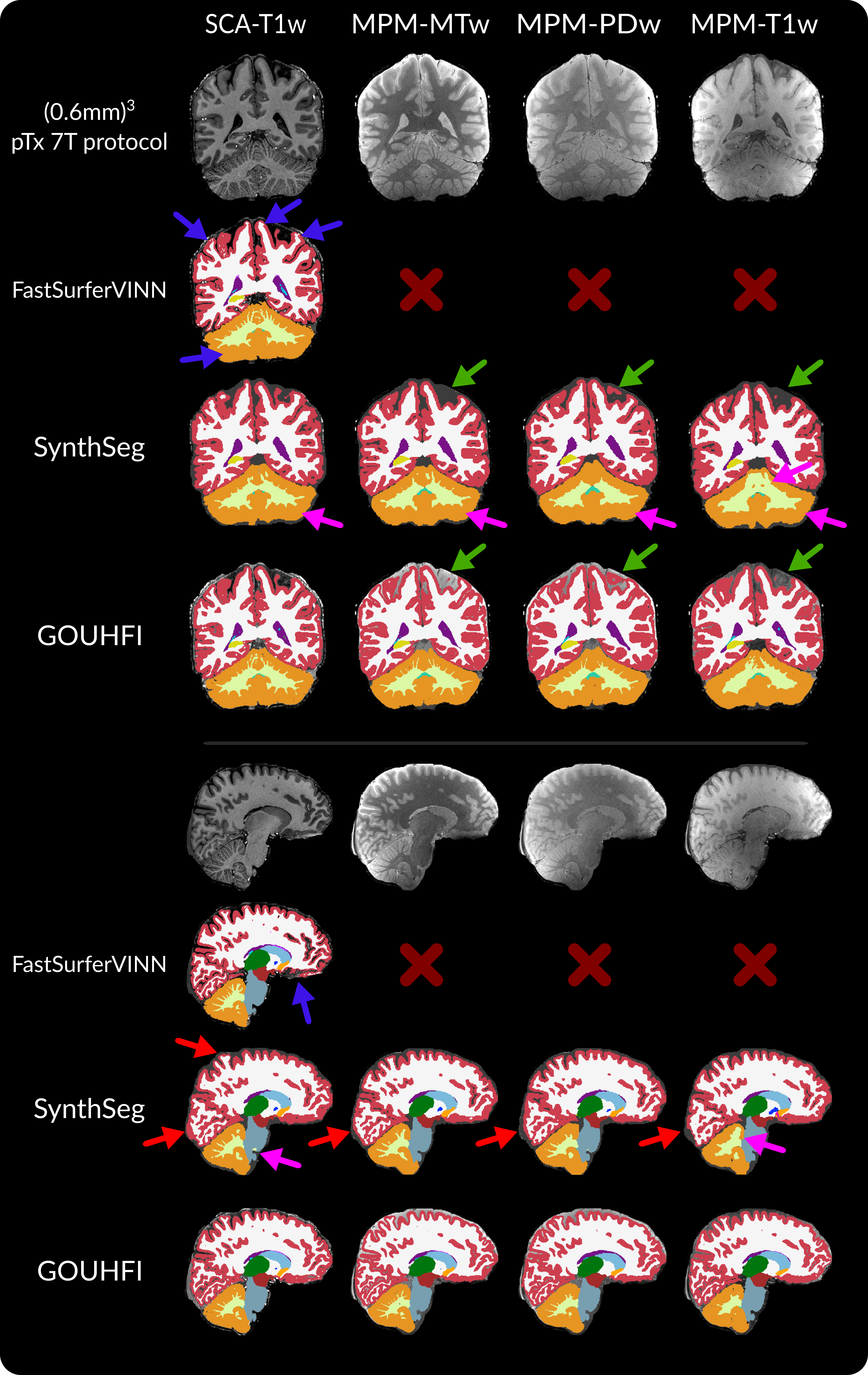}
        \caption{Segmentations produced by \fsv{} (first row), \sys{} (second row) and \go{} (bottom row) for one subject in the coronal (top part) and sagittal (bottom part) planes for the \tow{} \mpr{} (left), MPM-MTw (second from the left), MPM-PDw (second from the right) and MPM-T1w (right) contrasts from the SCAIFIELD dataset (7T). All images and segmentations have a resolution of \sixi{}. The \mpr{} images have been N4-corrected whereas all MPM contrasts have not. Segmentations from \fsv{} are shown for the \tow{} image only since it only segments \tow{} images. Blue arrows represent regions of mislabeling from \fsv{} (used as ground truth), whereas green arrows show discrepancies between the different MPM contrasts. Pink arrows represent mislabeling of cerebellum WM by \sys{}. Red arrows represent mislabeling between WM and cortex inside the cerebrum where \sys{} overestimated WM segmentation. The labels shown and their colors correspond to the \fss{} lookup table.}
        \label{fig:sca-segs}
    \end{center}
\end{figure}

In Figure \ref{fig:sca-segs}, the segmentations produced by \go{} and \sys{} for all four contrasts and by \fsv{} for the \tow{} \mpr{} are shown. As for the HCP dataset shown in section \ref{res:hcp}, the segmentations computed by \go{} and \sys{} are visually highly similar to the one from \fsv{} for the SCA-\tow{}. This is also demonstrated quantitatively with the DSC and ASD values reported in the first column of Table \ref{tab:dsc-asd-sca-mpm}. Even when using a pTx excitation combined with N4-correction for the \mpr{} images, signs of limited segmentation capacities for 7T images started to appear for \fsv{} as shown with the blue arrows in Figure \ref{fig:sca-segs}. Overall, \sys{} demonstrated a lower capacity to accurately delineate thin cerebellum WM branches in addition of grossly overestimating size of WM in small folded WM-cortex boundary regions across all contrasts tested compared to \go{} (red arrows on Figure \ref{fig:sca-segs}). For the SCA-\tow{} case, \go{} had consistently better DSC and ASD values than \sys{} for all labels tested with a marked difference for the cortex.

\begin{table}[h]
\centering
\begin{threeparttable}
\caption{Median DSC and ASD values (with 95\% CIs) computed for the four contrasts in the SCAIFIELD dataset (n=10) using \go{} and \sys{}. The ground truth for SCA-\tow{} was the segmentation produced by \fsv{} at native resolution using the N4-corrected pTx \tow{} images (GT\textsubscript{FSV}) whereas for the three MPM contrasts, the up-sampled segmentation produced by \sys{} was used (GT\textsubscript{\sys{}}). For the comparison of \go{} and \sys{} versus \fsv{}, the highest DSC (and lowest ASD) value is shown in bold.}
\scriptsize

\begin{tabular*}{\linewidth}{@{\extracolsep{\fill}}l c@{\hskip -1.5em}c c @{\hskip -1.5em}c @{\hskip -1.5em}c}
\toprule
\textbf{DSC} & \multicolumn{2}{c}{SCA-T1w (GT\textsubscript{FSV})} & \multicolumn{3}{c}{MPM (GT\textsubscript{\sys{}})} \\
            & \go{} & \sys{} & MPM-MTw$^*$ & MPM-T1w & MPM-PDw \\
\midrule
WM        & \textbf{0.97} [0.97, 0.97] & 0.92 [0.92, 0.92] & 0.92 [0.92, 0.93] & 0.90 [0.90, 0.90] & 0.90 [0.89, 0.90] \\
\addlinespace
Cortex        & \textbf{0.91} [0.90, 0.92] & 0.82 [0.81, 0.82] & 0.87 [0.87, 0.88] & 0.87 [0.86, 0.87] & 0.85 [0.84, 0.86] \\
\addlinespace
Putamen   & \textbf{0.94} [0.93, 0.94] & 0.91 [0.90, 0.91] & 0.91 [0.91, 0.92] & 0.88 [0.87, 0.89] & 0.87 [0.85, 0.88] \\
\addlinespace
Thalamus  & \textbf{0.94} [0.93, 0.94] & 0.92 [0.91, 0.92] & 0.91 [0.91, 0.93] & 0.93 [0.92, 0.93] & 0.92 [0.91, 0.92] \\
\addlinespace
Pallidum  & \textbf{0.88} [0.86, 0.89] & 0.86 [0.84, 0.86] & 0.85 [0.80, 0.87] & 0.80 [0.78, 0.22] & 0.78 [0.74, 0.81] \\
\addlinespace
Cerebellum WM   & \textbf{0.91} [0.90, 0.91] & 0.87 [0.87, 0.88] & 0.89 [0.88, 0.90] & 0.82 [0.81, 0.83] & 0.90 [0.89, 0.90] \\
\addlinespace
Cerebellum Cortex  & \textbf{0.94} [0.93, 0.94] & 0.91 [0.90, 0.91] & 0.93 [0.92, 0.93] & 0.92 [0.91, 0.92] & 0.93 [0.93, 0.93] \\
\addlinespace
\rowcolor{gray!10} 
\textbf{Median (27 labels)} & \textbf{0.91} [0.90, 0.91] & 0.88 [0.87, 0.88] & 0.89 [0.87, 0.89] & 0.87 [0.85, 0.86] & 0.84 [0.82, 0.84] \\
\bottomrule
\end{tabular*}

\vspace{1em}

\begin{tabular*}{\linewidth}{@{\extracolsep{\fill}}l c@{\hskip -1.5em}c c @{\hskip -1.5em}c @{\hskip -1.5em}c}
\toprule
\textbf{ASD [mm]} & \multicolumn{2}{c}{SCA-T1w (GT\textsubscript{FSV})} & \multicolumn{3}{c}{MPM (GT\textsubscript{\sys{}})} \\
            & \go{} & \sys{} & MPM-MTw$^*$ & MPM-T1w & MPM-PDw \\
\midrule
WM        & \textbf{0.13} [0.13, 0.14] & 0.36 [0.35, 0.36] & 0.44 [0.41, 0.46] & 0.54 [0.52, 0.57] & 0.57 [0.54, 0.61] \\
\addlinespace
Cortex        & \textbf{0.23} [0.22, 0.25] & 0.45 [0.44, 0.46] & 0.50 [0.47, 0.52] & 0.52 [0.49, 0.58] & 0.61 [0.59, 0.70] \\
\addlinespace
Putamen   & \textbf{0.26} [0.25, 0.31] & 0.47 [0.45, 0.50] & 0.41 [0.38, 0.44] & 0.54 [0.53, 0.65] & 0.63 [0.55, 0.67] \\
\addlinespace
Thalamus  & \textbf{0.41} [0.39, 0.48] & 0.49 [0.49, 0.55] & 0.55 [0.51, 0.65] & 0.50 [0.48, 0.55] & 0.61 [0.57, 0.66] \\
\addlinespace
Pallidum  & \textbf{0.43} [0.38, 0.46] & 0.45 [0.44, 0.51] & 0.72 [0.61, 0.87] & 0.78 [0.71, 0.82] & 0.79 [0.77, 1.01] \\
\addlinespace
Cerebellum WM & \textbf{0.32} [0.29, 0.34] & 0.45 [0.43, 0.51] & 0.51 [0.45, 0.54] & 0.69 [0.67, 0.77] & 0.53 [0.52, 0.58] \\
\addlinespace
Cerebellum Cortex & \textbf{0.49} [0.46, 0.55] & 0.65 [0.64, 0.67] & 0.59 [0.55, 0.63] & 0.65 [0.63, 0.69] & 0.65 [0.61, 0.65] \\
\addlinespace
\rowcolor{gray!10} 
\textbf{Median (27 labels)} & \textbf{0.34} [0.34, 0.38] & 0.48 [0.48, 0.52] & 0.50 [0.48, 0.53] & 0.56 [0.56, 0.61] & 0.64 [0.66, 0.72] \\
\bottomrule
\end{tabular*}

\begin{tablenotes}[flushleft]\footnotesize
\item[${*}$] A subset of 4 subjects was used for MPM-MTw since several subjects did not include a MTw MPM scan.
\end{tablenotes}

\label{tab:dsc-asd-sca-mpm}
\end{threeparttable}
\end{table}

\begin{figure}[htbp]
    \begin{center}
        \includegraphics[width=0.95\textwidth]{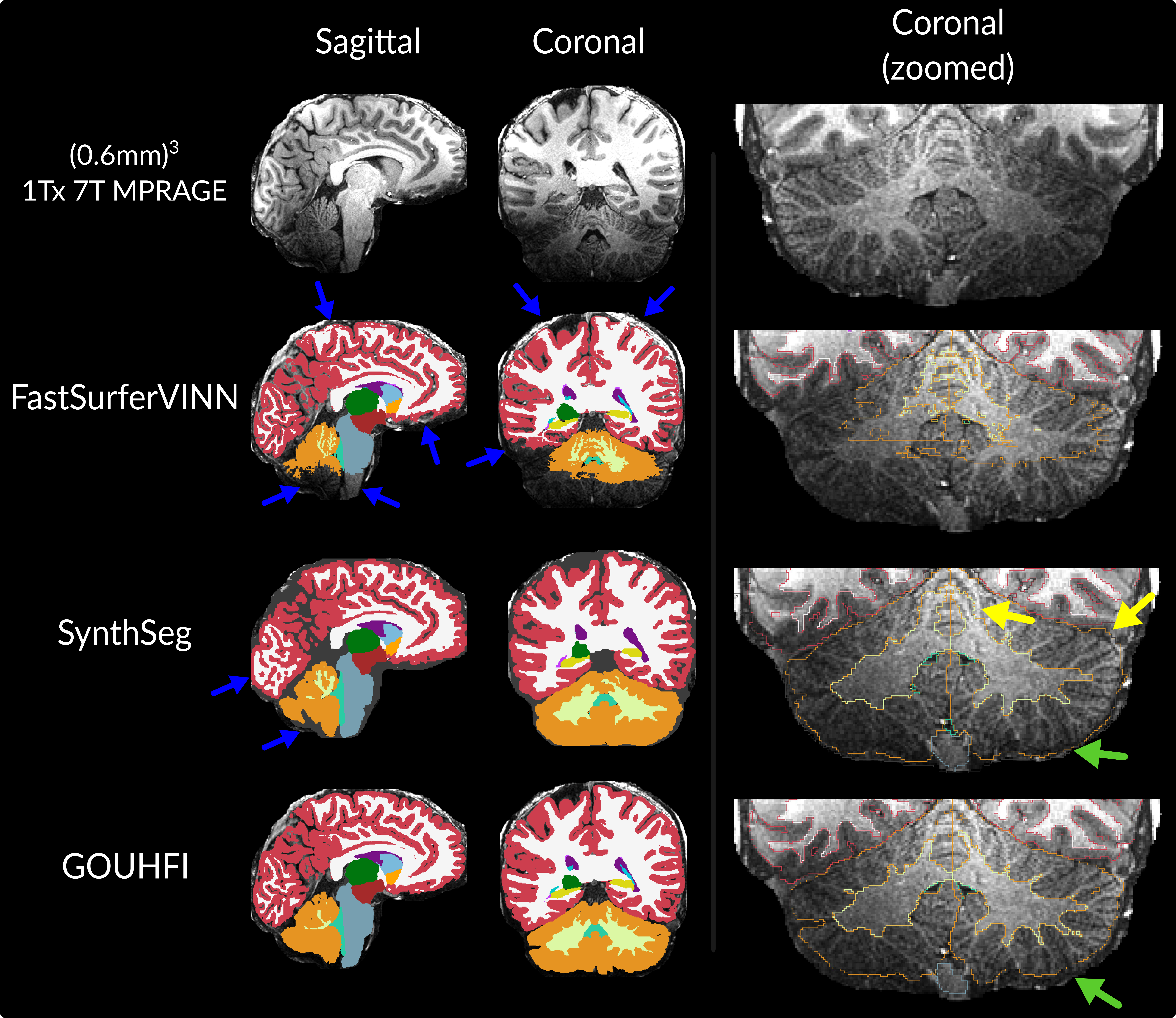}
        \caption{Visual comparison between the segmentations produced by \fsv{} (second row), \sys{} (third row) and \go{} (bottom row) on a 1 Tx \mpr{} acquired for one additional SCAIFIELD subject (7T). Significant signal and contrast inhomogeneities are present. This subject was neither included in the training nor in the testing datasets. The sagittal (first column) and coronal (second column) planes with a zoomed-in version of another coronal slice (third column) with the segmentation borders overlaid are shown. All images and segmentations have a resolution of \sixi{}. Blue arrows represent \fsv{} and \sys{} outputs being affected by signal inhomogeneities. The green arrows show the difference in cerebellar cortex delineation between \sys{} and \go{}. Yellow arrows show segmentation errors by \sys{} for the cerebellum WM and cortex. The labels shown and their colors correspond to the \fss{} lookup table.}
        \label{fig:sca-1tx}
    \end{center}
\end{figure}

The performance of \go{}, \sys{} and \fsv{} on one subject, for which an 1Tx \sixi{} \mpr{} acquisition was acquired (neither part of the training nor the test datasets) is demonstrated in Figure \ref{fig:sca-1tx}. \go{} and \sys{} created substantially better segmentations than \fsv{} as expected, especially in regions affected by signal and contrast alterations related to reduced RF transmit inhomogeneities. However, as similarly shown in Figure \ref{fig:sca-segs}, \sys{} also showed limited capacity to properly identify the boundary between WM and cortex in some regions compared to \go{}.

\subsection{Glasgow dataset: \go{} versus \ceb{}}

\begin{figure}[htbp]
    \begin{center}
        \includegraphics[width=0.75\textwidth]{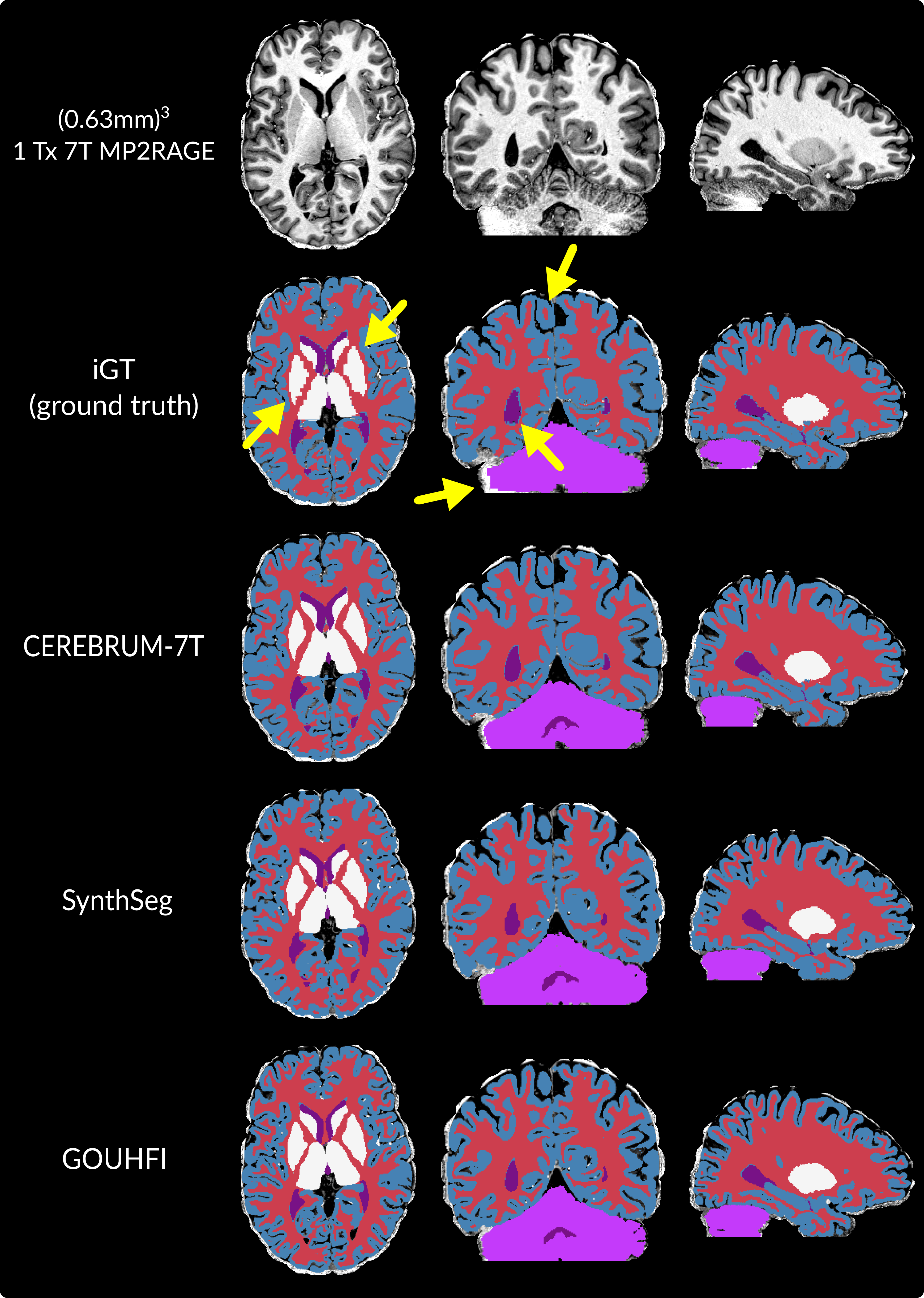}
        \caption{Segmentations produced by \ceb{} (third row), \sys{} (fourth row) \go{} (last row) with the corresponding iGT (second row the top) for one subject in all anatomical planes from the test dataset used for \ceb{}. All images and segmentations have a resolution of (0.63 mm)$^3$. Yellow arrows point to regions where the iGT (ground truth) seems sub-optimal compared to \ceb{}, \sys{} and \go{}. The labels shown here are gray matter (blue), white matter (red), ventricles (purple), basal ganglia (white) and cerebellum (violet). The brainstem is also segmented but not visible in this figure.}
        \label{fig:res-c7t-vs-gouhfi}
    \end{center}
\end{figure}

\go{} and \sys{} were tested against \ceb{}, the only brain segmentation technique optimized for 7T images. The results for one example subject are shown in Figure \ref{fig:res-c7t-vs-gouhfi}. The DSC and ASD values computed for \ceb{}, \sys{} and \go{} against the iGT are reported in Table \ref{tab:dsc-asd-c7t}. Both \go{} and \sys{} produced highly similar segmentations between each other and to \ceb{}, although \ceb{} being the method with the highest DSC and ASD with the iGT across all labels. 

\begin{table}[h]
\centering
\begin{threeparttable}
\caption{Median DSC and ASD values (with 95\% CIs) for each label across all test cases (n=21) using segmentations from \ceb{}, \go{}, and \sys{}. Ground truth is the iGT as described in \cite{svanera_cerebrum7t_2021}. The highest DSC (and lowest ASD) are shown in bold.}

\begin{tabular*}{\linewidth}{@{\extracolsep{\fill}}l c@{\hskip -1.5em}c@{\hskip -1.5em}c}
\toprule
\textbf{DSC} & \ceb{} & \go{} & \sys{} \\
\midrule
WM             & \textbf{0.94} [0.94, 0.94] & 0.92 [0.91, 0.92] & 0.90 [0.89, 0.91] \\
\addlinespace
Cortex         & \textbf{0.91} [0.90, 0.91] & 0.86 [0.86, 0.86] & 0.83 [0.83, 0.84] \\
\addlinespace
Basal ganglia  & \textbf{0.89} [0.89, 0.90] & 0.86 [0.86, 0.87] & 0.87 [0.86, 0.88] \\
\addlinespace
Ventricles     & \textbf{0.86} [0.85, 0.87] & 0.85 [0.83, 0.86] & 0.84 [0.83, 0.86] \\
\addlinespace
Brain-stem     & \textbf{0.93} [0.92, 0.93] & 0.91 [0.90, 0.91] & 0.91 [0.90, 0.91] \\
\addlinespace
Cerebellum     & \textbf{0.93} [0.88, 0.94] & 0.92 [0.86, 0.92] & 0.89 [0.83, 0.90] \\
\addlinespace
\textbf{Median} & \textbf{0.91} [0.90, 0.91] & 0.88 [0.87, 0.89] & 0.88 [0.86, 0.87] \\
\bottomrule
\end{tabular*}

\vspace{1em}

\begin{tabular*}{\linewidth}{@{\extracolsep{\fill}}l c@{\hskip -1.5em}c@{\hskip -1.5em}c}
\toprule
\textbf{ASD [mm]} & \ceb{} & \go{} & \sys{} \\
\midrule
WM             & \textbf{0.24} [0.24, 0.25] & 0.35 [0.34, 0.36] & 0.44 [0.43, 0.45] \\
\addlinespace
Cortex         & \textbf{0.28} [0.28, 0.29] & 0.43 [0.42, 0.44] & 0.50 [0.49, 0.50] \\
\addlinespace
Basal ganglia  & \textbf{0.48} [0.46, 0.50] & 0.60 [0.56, 0.63] & 0.59 [0.56, 0.63] \\
\addlinespace
Ventricles     & \textbf{0.35} [0.34, 0.45] & 0.39 [0.37, 0.46] & 0.38 [0.36, 0.47] \\
\addlinespace
Brain-stem     & \textbf{0.43} [0.41, 0.48] & 0.54 [0.51, 0.64] & 0.53 [0.50, 0.62] \\
\addlinespace
Cerebellum     & \textbf{0.68} [0.63, 1.21] & 0.90 [0.87, 1.51] & 1.26 [1.18, 1.89] \\
\addlinespace
\textbf{Median} & \textbf{0.38} [0.40, 0.52] & 0.49 [0.52, 0.66] & 0.51 [0.58, 0.76] \\
\bottomrule
\end{tabular*}

\label{tab:dsc-asd-c7t}
\end{threeparttable}
\end{table}

\subsection{MPI-CBS: \go{} and \sys{} performance for ultra-high resolution and inhomogeneous 7T images}\label{res:mpi}

\begin{figure}[htbp]
    \begin{center}
        \includegraphics[width=1\textwidth]{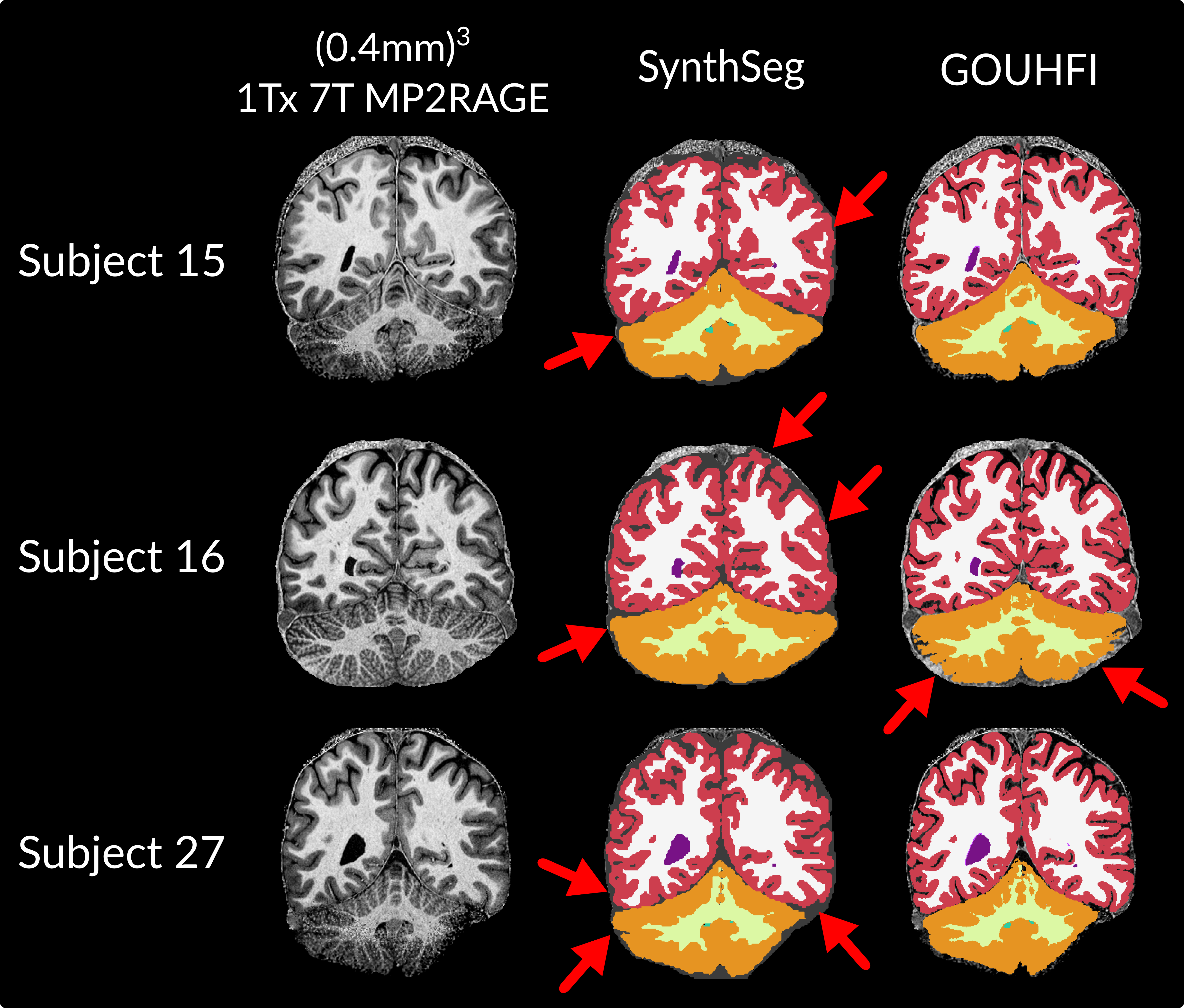}
        \caption{Segmentation results produced by \sys{} (middle column) and \go{} (right column) for three subjects from the MPI-CBS dataset in the coronal plane. All images were acquired at 7T with 1Tx \fori{} \mprr{} (same resolution for the segmentations). Red arrows point to segmentation errors in cortex and cerebellum cortex delineations.}
        \label{fig:quali-mpi-cbs}
    \end{center}
\end{figure}

Three example subjects from the MPI-CBS with \fori{} 1Tx \mprr{} acquired at 7T and their corresponding segmentations produced by \sys{} and \go{} are displayed in Figure \ref{fig:quali-mpi-cbs}. Although the network segments images at \sevi{} (resolution used by the network for training), the ultra-high resolution of this dataset posed no problem for \go{} to properly delineate the brain regions at \fori{}. However, since it was only trained with label maps at 1 mm$^3$, \sys{} showed an limited capacity to show the same level of details especially for the cortex and cerebellum WM branches as also reported in previous sections. Both techniques were able to manage the high level of inhomogeneity and noise present in the images. For subject 16, \sys{} showed superior identification of the cerebellum cortex in comparison with \go{}. However, in all cases, \sys{} systematically and inaccurately overextended laterally the cerebellum cortex. 

\subsection{UltraCortex: Performance of \go{} and \sys{} against manual white and gray matter delineations at 9.4T}

In Table \ref{tab:dsc-asd-uc}, the median DSC and ASD values computed for \go{} and \sys{} against the manual delineations for WM and GM are reported for the UltraCortex dataset. \go{} systematically outperformed \sys{} for every label and sub-dataset, with a substantial advantage for the cortex label with up to 7 Dice points improvement over \sys{}. 

\begin{table}[h]
\centering
\begin{threeparttable}
\caption{Median DSC and ASD values (with 95\% CIs) computed for the WM and GM segmentations (left and right hemispheres combined) produced by \go{} and \sys{} for subjects with manual segments provided in the UltraCortex dataset (\sixi{} \mprr{} n$_{sub}$=8, \sixi{} \mpr{} n$_{sub}$=3 and \eigi{} \mprr{} n$_{sub}$=1.) The highest DSC (and lowest ASD) are shown in bold.}
\footnotesize

\vspace{0.5em}
\begin{tabular*}{0.99\textwidth}{@{\extracolsep{\fill}}l *{3}{c@{\hspace{1em}}c}}
\toprule
\textbf{DSC} & \multicolumn{2}{c}{\sixi{} \mprr{}} & \multicolumn{2}{c}{\sixi{} \mpr{}} & \multicolumn{2}{c}{\eigi{} \mprr{}} \\
            & \go{} & \sys{}  & \go{} & \sys{}  & \go{} & \sys{} \\
\midrule
White Matter & \textbf{0.97} [0.97, 0.97] & 0.94 [0.94, 0.94] & \textbf{0.97} [0.96, 0.97] & 0.93 [0.93, 0.94] & \textbf{0.95} [-, -] & 0.93 [-, -] \\
Cortex       & \textbf{0.91} [0.91, 0.91] & 0.85 [0.84, 0.85] & \textbf{0.90} [0.89, 0.92] & 0.85 [0.84, 0.87] & \textbf{0.89} [-, -] & 0.83 [-, -] \\
\bottomrule
\end{tabular*}

\vspace{1.5em}

\begin{tabular*}{0.99\textwidth}{@{\extracolsep{\fill}}l *{3}{c@{\hspace{1em}}c}}
\toprule
\textbf{ASD [mm]} & \multicolumn{2}{c}{\sixi{} \mprr{}} & \multicolumn{2}{c}{\sixi{} \mpr{}} & \multicolumn{2}{c}{\eigi{} \mprr{}} \\
                 & \go{} & \sys{}  & \go{} & \sys{}  & \go{} & \sys{} \\
\midrule
White Matter & \textbf{0.26} [0.25, 0.29] & 0.39 [0.37, 0.39] & \textbf{0.27} [0.25, 0.32] & 0.44 [0.43, 0.44] & \textbf{0.35} [-, -] & 0.44 [-, -] \\
Cortex       & \textbf{0.25} [0.25, 0.27] & 0.45 [0.44, 0.46] & \textbf{0.29} [0.26, 0.35] & 0.44 [0.39, 0.47] & \textbf{0.33} [-, -] & 0.48 [-, -] \\
\bottomrule
\end{tabular*}

\label{tab:dsc-asd-uc}
\end{threeparttable}
\end{table}

\subsection{STRAT-PARK: Parkinson's disease volumetry study at 7T}

\begin{figure}[htbp]
    \begin{center}
        \includegraphics[width=1\textwidth]{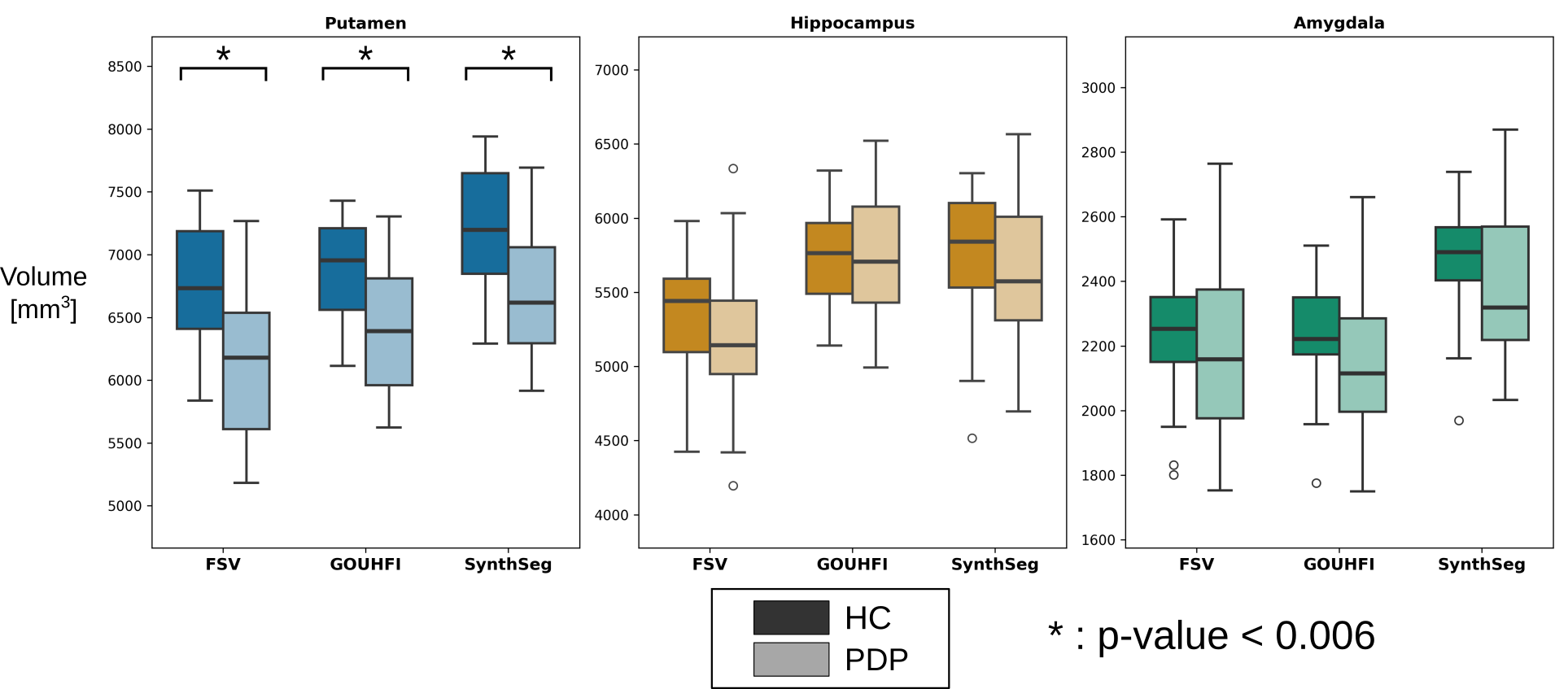}
        \caption{Box plots showing the normalized volumes measured by \fsv{} (left), \go{} (middle) and \sys{} (right) for healthy controls (HC) and Parkinson's disease patients (PDP) for the putamen, hippocampus and amygdala. For putamen, the three techniques had a statistically significant difference in volume between HC and PDP after Bonferroni correction.}
        \label{fig:boxp-volum}
    \end{center}
\end{figure}

The volumetric analysis results are shown in Figure \ref{fig:boxp-volum}. The same consistent decrease trend between HC and PDP was observed for all techniques for the putamen, hippocampus and amygdala. It was only for putamen that all techniques presented a statistically significant difference between both HC and PDP sub-groups. For \go{}, the median volumes measured were larger than \fsv{} for both HC and PDP whereas the opposite trend was observed for the amygdala. On the other hand, the median volume computed for the amygdala for \sys{} was considerably and unexpectedly larger than both \fsv{} and \go{}. The p-values calculated were the following for \fsv{}/\go{}/\sys{}: 0.002 / 0.004 / 0.0004 for putamen, 0.22 / 1.0 / 0.31 for hippocampus and 0.36 / 0.06 / 0.27 for amygdala (Bonferroni-correct significance threshold: p-value $<$ 0.006).

\subsection{Human Brain Atlas: Ultra-High Resolution at 7T} 

\begin{figure}[htbp]
    \begin{center}
        \includegraphics[width=0.90\textwidth]{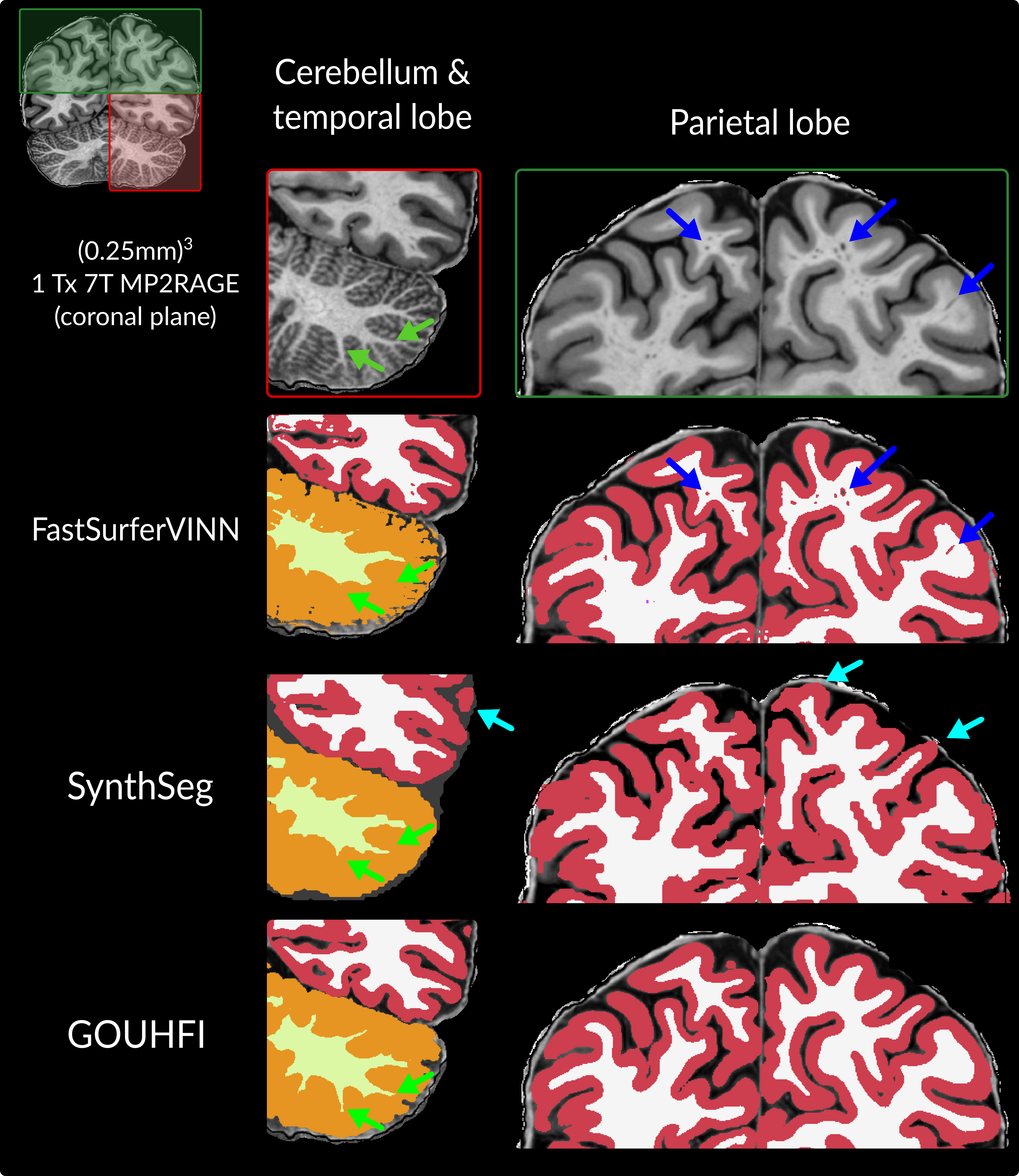}
        \caption{Segmentations produced by \fsv{} (second row), \sys{} (third row) and \go{} (last row) at (0.25mm)$^3$ for the averaged \tow{} image in the coronal plane for subject 001 from the Human Brain Atlas dataset (7T). The first column shows a zoomed-in version of the cerebellum and temporal lobe whereas the second column shows the parietal lobe. Green arrows show differences in segmentations between the three methods for fine cerebellar WM branches and their corresponding segmentations whereas blue arrows show the perivascular spaces inside WM. Turquoise arrows point to cortex segmentation errors for \sys{}.}
        \label{fig:hba-025}
    \end{center}
\end{figure}

Zoomed-in coronal views of the \twohi{} segmentations produced by \fsv{}, \sys{} and \go{} for the cerebellum and parietal lobe are presented in Figure \ref{fig:hba-025}. For \fsv{}, a significant amount of cerebellar WM branches were not segmented even in regions not affected by signal inhomogeneities as shown with the green arrows. Although improvements were noticeable with \sys{}, the best overall detection and segmentation of cerebellum WM was done by \go{}. Moreover, perivascular spaces (PVS) in WM, which become more easily visible at this resolution, were often segmented as background or cortex for \fsv{} (blue arrows in Figure \ref{fig:hba-025}) whereas \sys{} and \go{} segmented them as WM. \sys{} showed limitations in some cortex regions with noticeable mislabeling of non-cortical voxels as cortex. Furthermore, \sys{} segmentation showed non-smooth, step-like delineations, which \fsv{} and \go{} did not show.

%

\subsection{Impact of label granularity from 3T to UHF-MRI} 

\begin{figure}[htbp]
    \begin{center}
        \includegraphics[width=0.95\textwidth]{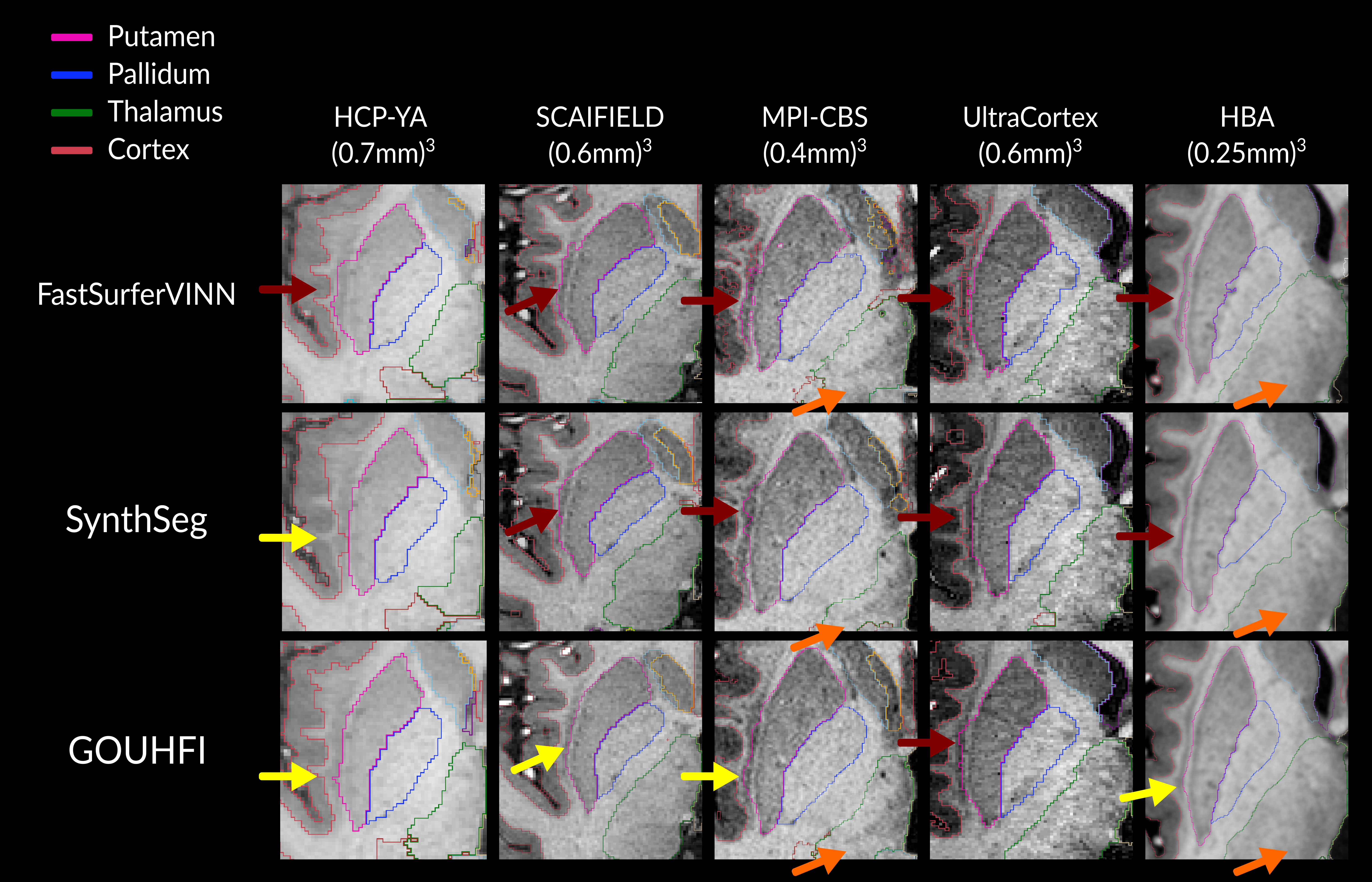}
        \caption{Close-up view of an axial slice showing the segmentations of the putamen, pallidum, thalamus and cortex in the right hemisphere by \fsv{} (first row), \sys{} (second row) and \go{} (last row) overlaid on the corresponding \tow{} image used for segmentation. Red arrows show cases where significant parts of the claustrum are segmented as putamen. Yellow arrows show cases where a small portion of the claustrum is included or that the boundary of the putamen is slightly misaligned with its actual border while not including the claustrum. Orange arrows represent ultra-high resolution cases where subfields of the thalamus can be observed while not being properly segmented.}
        \label{fig:claus-new}
    \end{center}
\end{figure}

Figure \ref{fig:claus-new} illustrates the impact of training \go{} with label maps generated by \fsv{}, tailored to the granularity level typical at 1.5–3T, when applied to UHF images. \fsv{} performed the poorest among the three techniques at properly delineating the putamen by systematically including the claustrum in all examples shown. On the other hand, although some small portions of the claustrum were still included or, alternatively, the boundary of the putamen was slightly misaligned, \go{} performed the best and was the least affected technique by this systematic error among the three. Additionally, for the MPI-CBS and HBA cases where different sub-fields boundaries of the thalamus were discernible, all techniques struggled to accurately identify the thalamus boundary, which resulted in all of them creating a fictitious boundary that did not reflect the internal contrast observed. 

\section{Discussion}\label{disc}

In this study, a novel DL-based segmentation technique capable of segmenting brain MR images of any contrast and resolution is proposed. As shown, \go{} was able to accurately segment MR images acquired at 3T, 7T and 9.4T with a total of six different contrasts and seven different resolutions. \go{} performed well on highly inhomogeneous 1Tx images acquired at 7T and 9.4T where standard tools are prone to failure. Moreover, \go{} demonstrated highly similar performance against domain-specific techniques like \fsv{} or \ceb{} when tested in their respective domains while also consistently outperforming \sys{}, the only DL-based contrast-agnostic segmentation tool available. Ultimately, when used to assess its ability to detect volumetric changes in Parkinson's disease, \go{} showed similar performance to \fsv{} and \sys{} to monitor volume losses in accordance with the literature. 

\subsection{Training label maps}

All previous DL-based segmentation techniques that have been trained on automatically produced label maps have used \fss{} to produce the training label maps. To the best of our knowledge, \go{} is the first technique to use \fsv{-based} label maps in the training dataset. This choice was made based on the fact that the label maps produced by \fss{}, even with the sub-millimeter option selected, produced coarse delineations of the labels (see Figure \ref{fig:appC} in Appendix \ref{appC}). While, in theory, the label maps created by \fss{} are at the same resolution as the input images, the "effective resolution" of the label maps is visibly lower. Considering that only sub-millimeter resolutions were used for training and that the intended usage of \go{} is for sub-millimeter images, \fsv{} was preferred since it produced more refined delineations than \fss{}.

In this study, only automatically produced label maps were used for the training corpus, as it was done for \fsv{}. While this could be a potential issue when using real UHF images for training, this is not the case when using synthetic images. If minor segmentation errors with respect to the corresponding real \tow{} input image would be present in the label map (e.g., small parts missing from the cerebellum or temporal lobes due to inhomogeneities), these errors would be "lost" during the creation of the synthetic image for that subject. By design, the synthetic images and corresponding label maps are perfectly aligned with each other. As mentioned in \cite{billot2023synthseg}, and again demonstrated in this study, the usage of automatically produced label maps for synthetic training data is not only possible, but highly recommended since it allows to considerably increase the number of training cases. In fact, even for techniques like \fsv{}, which are not based on synthetic training images, the size of the training corpus was shown to be the most important factor to improve the model \citep{henschel_fastsurfervinn_2022}. Ultimately, as mentioned in the methods section, extensive visual QA was done on the label maps produced by \fsv{} for the UHF images before including them in the study (only 15 out of the 78 subjects were kept from the UltraCortex dataset). Since real images were used for validation, we had to make sure that both the quality of the \tow{} images and label maps were good enough since mismatch between both would negatively impact the validation process.

\subsection{HCP-YA: Benchmarking against \fsv{} \& \sys{} at 3T}

For both contrasts tested from the 3T HCP-YA dataset, \go{} performed remarkably well and systematically better than \sys{}. The only ROIs where \sys{} produced better DSC and ASD values than \go{} were for the cerebellar WM and cortex. That can be explained by the similar behavior of \sys{} to reproduce the limited identification of inferior cerebellar WM branches like \fsv{} does. Indeed, \go{} detected substantially more cerebellar WM than both \fsv{} (reference technique) and \sys{}, resulting in lower DSC and ASD for both cerebellar WM and cortex. Interestingly, \go{} used the same DR approach to create the synthetic training data as \sys{}. However, for \go{}, it resulted in a superior detection of the thin cerebellar WM branches and cortex sulci. Thus, the fact that (1) \go{} was trained using only sub-millimeter label maps (with \sevi{} as the training resolution) and (2) the randomized downsampling step as done in \sys{} was disabled could explain this improved identification of high-resolution anatomy features. 

Moreover, the \tow{} \mpr{} images gave higher DSC and lower ASD values than the \ttw{} images, although only \ttw{} from this dataset was used for the validation set during training. One can argue that \tow{} was well represented in the validation dataset. However, all other \tow{} images used for validation were either from a different vendor (ABIDE-II ETH: Philips Achieva with 3D TFE sequence \& ABIDE-II EMC: GE MR750 with IR-FSPGR sequence) or different field strength and sequence (UltraCortex: 9.4T with \mprr{}). Another possible explanation for this could be that it is a consequence of using label maps originally produced from \tow{} images for the creation of the synthetic training data, which all exhibit the same \tow{}-visible structures. Ultimately, the finding that \go{} produced segmentations with DSC $\geq$0.88 and ASD smaller than one voxel over 27 labels and 20 subjects in \fsv{'s} domain (i.e., 3T \sevi{} \tow{} \mpr{}) is a strong indication of its robustness and comparable performance for segmentation tasks even outside \go{}'s optimized domain (i.e., UHF-MRI) while also being superior to alternatives like \sys{}.

\subsection{SCAIFIELD: Contrast- and resolution-agnostic performance at 7T}

The SCAIFIELD dataset served as an excellent test dataset for \go{} considering its variety of contrasts, resolution different from the trained one, and its UHF nature. \go{} demonstrated its contrast-agnostic performance by segmenting all four contrasts well. Overall, \go{} showed a significantly higher level of details than \sys{}, especially within the cortex and both the cerebellar WM and cortex labels as similarly reported for the HCP-YA dataset. These observations reinforce the idea that \sys{'s} use of \onei{} training resolution, combined with the random down-sampling of the training label maps, negatively impacts the quality of the segmentations for UHF images.

For the three MPM contrasts, with \sys{} as the reference technique, the MPM-PDw image appeared to be the most challenging to segment based on the quantitative metrics computed. However, one could argue that the segmentations displayed in Figure \ref{fig:sca-segs} for \go{} showed a lower level of detail for the MPM-\tow{} compared to the MPM-PDw, especially in the cerebellum. Ultimately, the better quantitative performance of MPM-\tow{} over MPM-PDw might be explainable by one simple observation: \sys{} has limited capacity to segment lower contrast regimes at high-resolutions like the MPM-PDw and MPM-\tow{}, resulting in a poor reference to compare \go{} with. Nonetheless, it is important to mention that \go{} was also challenged by the low level of contrast in the MPM-\tow{}, but appeared to segment only visible structures rather than inferring or "hallucinating" invisible regions (cf. pink arrows pointing to cerebellum WM in the coronal view of \sys{} for the MPM-\tow{}). Ultimately, the "better" quantitative performance for the MPM-\tow{} is probably due to a poorer but "matched" performance between \sys{} and \go{}.

Similar to the \ttw{} versus \tow{} contrast-agnostic comparison using the HCP-YA data, it is interesting to observe that the MPM-PDw dataset showed the lowest quantitative agreement with the reference, while at the same time being the contrast used for the validation dataset for SCAIFIELD data during training. 

While being contrast-agnostic is a great feature of \go{} (and \sys{}), one inherent aspect of this is its strong generalization to \tow{} contrast variations. \tow{} contrast is considered the standard for high-resolution anatomical brain images, however, there is still a wide variety of implementations of \tow{} contrasts. Indeed, whether it is a difference rising from sequence selection (e.g., \mpr{} versus \mprr{}), choice of acquisition parameters (e.g., TI, TE and FA values) or even vendor implementations (e.g., Siemens' \mpr{} versus GE IR-FSPGR), \tow{} can appear quite different across centers or neuroimaging studies. Therefore, as shown in \cite{billot2023synthseg}, DL segmentation techniques trained on specific \tow{} images showed poor generalization to other \tow{} contrasts, whereas contrast-agnostic techniques like \sys{} and \go{} performed remarkably well and even better in some cases. Essentially, even if not designed or optimized for \tow{} contrast, \go{} should still be considered as a robust and accurate segmentation option for \tow{} datasets.

For the SCA-\tow{}, \fsv{} segmentations were still chosen as ground truth over \sys{} (used for the three MPM contrasts). From ad hoc qualitative assessment of the segmentation quality, \fsv{} was deemed a superior segmentation technique over \sys{} even if it was not designed for 7T and $<$ \sevi{} images. Previous work \citep{fortin2025} showed that \fsv{} performed quite well when N4-correction and pTx pulses were used. Indeed, as seen from Figure \ref{fig:sca-segs}, the \mpr{} image does not exhibit the typical strong signal inhomogeneities observed at 7T, and the full cerebellum and temporal lobes were properly detected by \fsv{}. \sys{} showed poor boundary detection between WM and cortex in highly gyrified regions (red arrows on sagittal view of Figure \ref{fig:sca-segs}) probably due to its low training resolution (+9 Dice points for the cortex label for \go{} compared to \sys{} [0.91 versus 0.82 respectively]). This was determinant in the decision to not pick \sys{} as the ground truth for the SCA-\tow{} case. On the other hand, signs of limitations for \fsv{} could be observed such as cortical voxels being mislabeled as WM in some cases, or a significant number of cerebellar WM branches not being detected (common issue with \sys{}). It is also important to mention that the latter issue did not seem to be a 7T-specific issue as cerebellar WM branches appeared to be also difficult to segment at 3T for the \tow{} images from the HCP dataset (see Figure \ref{fig:hcp-segs}). 

Moreover, it is essential to highlight that even if the theoretical best DSC score achievable is 1, in this work, DSC scores between 0.85-0.90 were desired due to the use of a "silver standard" as ground truth. For most test cases tested in this work where \fsv{} was set as the reference, it must be highlighted that authors were fully aware that it was not expected to perform well outside the HCP and pTx SCAIFIELD test scenarios. For non-\tow{} contrasts, \sys{} was the best and only DL contrast-agnostic technique available to compare \go{} with. However, its low training resolution applied to high-resolution images made it a questionable choice as a reference as discussed previously in this section. Thus, in the SCAIFIELD MPM case, the DSC and ASD scores reported against \sys{} should not be interpreted as direct quantitative assessment of \go{'s} performance for UHF-MRI, but rather as a general indicator of how it compared to \sys{}. Ultimately, this further emphasizes the necessity for novel segmentation techniques to be developed for UHF-MRI.

In all cases tested in this work, \go{} demonstrated improved segmentation of cerebellar WM branches over \fsv{} and \sys{}, even in cases like HCP-\tow{} where both should be expected to be superior. This makes \go{} particularly interesting for neuroimaging studies where the cerebellum is of importance, like for spinocerebellar ataxias \citep{ferreira2024cerebellar, arruda2020volumetric}, whether it is UHF-MRI or not.

Results displayed in Figure \ref{fig:sca-1tx} showcased \go{'s} and \sys{'s} capacity to segment highly inhomogeneous UHF images, significantly better than \fsv{}. Even in the cerebellum region with extremely low signal, \go{} was able to properly delineate the cerebellar GM and WM whereas \sys{} "over-segmented" the superior region of the cerebellar WM in a similar fashion as for the MPM-\tow{} as previously discussed. Conversely, even for cortical regions in the parietal and frontal lobes affected by hyperintense signal, \go{} accurately detected WM and GM voxels. \sys{} showed limitations in properly delineating the fine cortical regions with frequent mislabeling resulting in overly segmenting non-cortical voxels as cortex. Overall, \sys{} showed the same level of resistance to signal inhomogeneity as \go{} with most of its limitations probably due to the low training resolution. Since a substantial level of noise was present in the inferior part of the cerebellum, it was not clear where the actual border of the cerebellum was. It is fair to say that \sys{} proposed a more cautious estimate of it compared to \go{}. However, \sys{} has repetitively shown signs of over-cautiousness with this results (yellow arrows on Figure \ref{fig:sca-1tx}) and with the MPI-CBS dataset too. Nonetheless, the best identification of the cerebellum GM between \sys{} and \go{} is quite challenging to assess with certainty. Ultimately, the increased random signal bias implemented in the generative model used by \go{} for the creation of synthetic training images did not seem to meaningfully modify the overall high inhomogeneity-resistance that was already present with \sys{'s} generative model.  

As a result, this resistance of \go{} to high levels of noise, granularity and inhomogeneity is a direct outcome of the use of synthetic images for training. As shown with the example dataset in Figure \ref{fig:pipe}, the synthetic images exhibited similar features to typical UHF images due to the randomly simulated noise and inhomogeneities generated in the images while the corresponding segmentations remained unaffected. This would not be possible if real images were used for training, since the segmentations would be directly affected by the noise and inhomogeneity levels present in the input images. 

\subsection{Glasgow dataset: \go{} versus \ceb{}}\label{disc:ceb}

Both \go{} and \sys{} performed as well against \ceb{} and its iGT. Even if the quantitative metrics were slightly lower for \go{} and \sys{} compared to \ceb{}, we would like to argue that this might be the consequence of using the iGT as the ground truth. Indeed, suboptimal delineations were observable in the iGT segments as shown in Figure \ref{fig:res-c7t-vs-gouhfi} with the yellow arrows. For instance, coarse delineations were present especially for small cortical regions or the basal ganglia. Moreover, \ceb{} falsely assigned voxels affected by partial volume effects at the border of ventricles and WM as gray matter, which was also the case for the iGT but not \go{} nor \sys{}. Moreover, the cerebellum segmentation in the iGT was noticeably poor.

Since \ceb{} is the only segmentation technique optimized for 7T images, one might argue that it should have been the preferred technique for comparison against \go{} in this study. However, two main reasons can explain why \sys{} (or even \fsv{}) was preferred. First, \ceb{} only segments the brain into six labels. For instance, \ceb{} only generates one label for the basal ganglia. This considerably limits the usability of \ceb{} in neuroimaging studies where the individual subcortical nuclei like the thalamus or putamen can be of interest \citep{solomon2017mri,rua2020multi}. The same argument applies for the amygdala and hippocampus which were unusually combined with the rest of GM into one single label. Finally, the segmentation of the cerebellum as one label, that does not differentiate between cerebellar WM and GM, is also limiting. At 3T and UHF-MRI, these structures are even frequently segmented into smaller sub-nuclei for more precise analyses \citep{haast2024insights, keuken2013ultra, plantinga2018individualized, faber_cerebnet_2022, morell2024deepceres}. Considering that \ceb{} used \fss{} \textit{v6} to obtain individual subcortical nuclei segments to then recombine them under the basal ganglia as one single segment, it raises the question of why this unusual choice, especially for an UHF-MRI dedicated tool, was made.

The second issue related to \ceb{} was the technical prerequisites in order to use it. Out-of-the-box, \ceb{} can only segment images respecting these four requirements: 1) (0.63mm)$^3$ resolution; 2) \mprr{} sequence; 3) matrix size of 256$\times$352$\times$224; and 4) images from the Glasgow dataset. Any divergence from one of these four requirements requires fine-tuning \citep{svanera_cerebrum7t_2021}. For instance, as discussed in \cite{henschel_fastsurfervinn_2022}, no de facto standard resolution exists for high-resolution images, and even less at UHF-MRI, making the first requirement quite constraining in a similar fashion as it is for \sys{} with only \onei{} outputs. While a fine-tuning process requires less time and data than a full DL training, the user still faces practical challenges similar to a full training (i.e., having access to considerable GPU hardware with a significant amount of data curation and preparation) in addition to the prerequisite of a few, already available, high quality segmentations from their specific dataset. The latter can be interpreted as a circular dependency, where segmentations are actually required in order to produce segmentations. Therefore, in this work, fine-tuning \ceb{} to every test dataset including 7T data was considered unfeasible due to the complexity of producing high quality segmentations for these datasets, which exceeded the realistic scope of this work. Moreover, implementing a segmentation technique with these requirements in a clinical setting would be extremely challenging. 

Ultimately, considering all practical challenges related to the usage of \ceb{} on unseen data and its modest number of labels segmented, using \ceb{} at UHF-MRI is considerably limiting and thus explains its absence for other test datasets in this work.

\subsection{MPI-CBS: \go{} and \sys{} performance for ultra-high resolution and inhomogeneous 7T images}

Even when sequences like \mprr{} are used to reduce the impact of inhomogeneities with 1Tx at 7T, regions like the cerebellum are frequently affected by poor contrast-to-noise ratios, as shown by Figure \ref{fig:quali-mpi-cbs}. Both \sys{} and \go{} performed remarkably well to identify the full cerebellum, with actually a superior identification of the inferior border of the cerebellum by \sys{} for subject 16. However, \sys{} systematically and erroneously segmented subarachnoid spaces on both sides of the cerebellum as cerebellum GM for all subjects shown. Additionally, for subject 27, \sys{} struggled to properly identify and delineate the cortex inside the temporal lobe for both hemispheres.

In the end, this test dataset demonstrated the clear limitation of inferring with a model that was trained with a lower resolution (\onei{} or even lower considering the random down-sampling) in terms of properly segmenting fine gyrified cortex regions at ultra-high-resolution like \fori{}. That resulted in frequent mislabeling of the cortex with frequent poor overestimation of the extend of its actual localization (red arrows on Figure \ref{fig:quali-mpi-cbs}).   

\subsection{UltraCortex: Performance of \go{} and \sys{} against manual white and gray matter delineations at 9.4T}

Both \go{} and \sys{} showed great accuracy when compared to manual WM and GM segmentations with DSC $>$0.89 for all three sub-datasets from the UltraCortex 9.4T dataset. However, \go{} was consistently superior to \sys{} across all sub-datasets, with marked superiority for the cortex segmentation with at least 6 points of improvement on the median DSC. As already discussed in previous sub-sections, this is another example of the limited capacity of \sys{} to properly segment highly gyrified cortex regions compared to \go{}. However, this time, the ground truth is the gold standard with manual delineations, which gives even greater credibility to this observation. After extensive search, this dataset was found to be the only dataset with manual segmentations of complete ROIs available for sub-millimeter images acquired at UHF-MRI. It would have been highly interesting to evaluate \go{} against manually segmented subcortical structures at their native sub-millimeter resolution, but we were unable to find such dataset, presumably due to the extensive amount of time and expertise required to execute such a task.

\subsection{STRAT-PARK: Parkinson’s disease volumetry study at 7T}

\fsv{}, \go{} and \sys{} were able to detect volumetric changes between HC and PDP as shown in Figure \ref{fig:boxp-volum}. The consistent decrease in volumes between HC and PDP is in agreement with the literature for these three ROIs \citep{junque2005amygdalar,pieperhoff2022regional,geng2006magnetic}.

However, one difference observed between \fsv{} and both \go{} and \sys{} was the larger median hippocampal volume computed for both HC and PDP compared to \fsv{} (middle plot in Figure \ref{fig:boxp-volum}). Ad hoc qualitative observations of the segmentation results indicated that, in certain subjects with substantially enlarged ventricles, \go{} tended to overestimate hippocampal volume by erroneously including portions of the adjacent inferior lateral ventricle inside the hippocampal delineation. That was not observed for \fsv{} nor \sys{}. Nonetheless, this hippocampal over-segmentation did not appear to impact the segmentation quality for the amygdala for \go{}. This highlights a potential limitation of the generative model used by \go{}, namely its inability to synthesize unhealthy brain anatomies where subtle anatomical deviations from healthy brains can impact its performance. As for any automated segmentation technique, we recommend the users to visually inspect their segmentation results. Incorporating a more diverse training dataset with older subjects could possibly help mitigating this issue. Ultimately, further clinical validation and analyses on more diverse and aged clinical cohorts with different neurological conditions should be done for \go{} in the future, but is currently outside the scope of this work.

\subsection{Human Brain Atlas: Ultra-High Resolution at 7T}

Reaching ultra-high resolution levels like for the (0.25mm)$^3$ \mprr{} images from the Human Brain Atlas dataset allows for the visualization of PVS. Indeed, PVSs in healthy subjects have diameters between 0.13 mm to 0.96 mm, with the majority being below 0.5 mm \citep{zong2016visualization}. \go{} and \sys{} did not segment any PVS (it was included in WM) whereas \fsv{} segmented most of them and labeled them as background or cortex. Whether PVS should be segmented as part of WM is subject to discussion. However, segmenting them as cortex is incorrect. Especially at UHF-MRI, researchers should start considering specific inclusion of PVS in label maps, in particular as the number of studies about PVS has drastically increased in the recent years \citep{kilsdonk2015perivascular, george2021novel, feldman2018quantification, wardlaw2020perivascular}. 

The (0.25mm)$^3$ segmentations produced by \go{} showed great delineation of the structures despite the fact that this resolution was considerably outside the training resolution of \sevi{}. Conversely, \sys{} did not show the same level of delineation for the segmentations, even if the same up-sampling approach as \go{} was used. These results further demonstrated the clear improvement in delineation quality by using a higher training resolution for \go{} over \sys{}. In Figure \ref{fig:hba-025}, one can observe the cortex segmentation created by \sys{}, even if up-sampled to (0.25mm)$^3$, has jagged contours and several regions of overextending the cortex into the bordering CSF (turquoise arrows). Both these behaviors were not observed for \go{}. The substantial difference in voxel volume between the output resolution of \sys{} and the input HBA image (64 times bigger voxels) made the jagged contours more apparent, although such artifacts were also present, albeit less visibly, at other sub-millimeter resolutions.  

Indeed, \go{} (and \sys{}) uses an "\textit{external scaling}" (exSA) approach to deal with any resolution instead of the "\textit{internal scaling}" (or VINN) approach as proposed in \fsv{}. While in \cite{henschel_fastsurfervinn_2022} the results were consistently better for the VINN approach over the exSA approach for all datasets shown in their Figure 8 (and the corresponding Table 9), none of the datasets showed a significantly better performance statistically, with only minimal improvement of the mean values compared to the standard deviations. In fact, for the subcortical structures of the ADNI dataset, the DSC and ASD values reported in Table 9 were actually higher for exSA over the VINN approach, which is in disagreement with the results reported in their Figure 8. Overall, the differences reported between the exSA and VINN approaches were larger for cortical than subcortical structures (i.e., only two labels, left- and right-cortex, compared to the 33 other labels in this work). 

One technical challenge of using the VINN technique (like \fsv{}) is the substantial increase in required GPU hardware to segment ultra-high resolutions, like the (0.25mm)$^3$ used here, since the full matrix is fed to the network. For instance, the GPU hardware used in this work was not powerful enough to segment the (0.25mm)$^3$ images due to VRAM limitations. This resulted in the required usage of CPU resources to segment the (0.25mm)$^3$ images using \fsv{}, which increased computation time by a factor of 78. While this was not an issue for a single subject dataset, using the VINN approach might be limiting for the increased matrix size of images with ultra-high resolutions like the HBA case and is something to consider for larger datasets or researchers with limited access to GPU hardware.  

\subsection{Impact of label granularity from 3T to UHF-MRI} 

As reported in \cite{valabregue2024comprehensive}, \fss{} and \fsv{} have been shown to erroneously include a substantial portion of the claustrum within the putamen segmentation. Since both \sys{} and \go{} were trained using label maps produced by \fss{} and \fsv{} respectively, the same pattern should have been expected especially due to the higher contrast at UHF. Surprisingly, as demonstrated with Figure \ref{fig:claus-new}, \go{}, while not being able to perfectly delineate the putamen, did not exhibit as poor delineations of the putamen as \fsv{}, and overall better than \sys{}. Nevertheless, when tested on ultra-high resolution and high \tow{} contrast like for the MPI-CBS and UltraCortex datasets, \go{} performed suboptimally in a similar fashion as \sys{}. One additional related issue, specific to \sys{}, is its poor delineation of the cerebral cortex which frequently resulted in the cortex and putamen being directly segmented next to each other (cf. yellow arrow on the HCP-YA example). 

Moreover, an issue arising from using label maps defined at the 3T-granularity level is the absence of sub-field distinction for some subcortical nuclei. Given that UHF-MRI offers increased resolution and contrast, this can become a problem for some structures like the thalamus, hippocampus or amygdala. Indeed, especially for the MPI-CBS and HBA cases with ultra-high resolution at 7T, different contrasts were visible and present inside the thalamus label. While this single thalamus label definition was adequate at 1.5 and 3T, this definition becomes limiting in some instances at 7T with ultra-high resolution and contrast as shown here. 

Ultimately, the limitations observed for both the claustrum and thalamus in \fsv{}, \sys{} and \go{} underscore the need for label definitions adapted to the granularity of UHF images. This adaptation, not widely implemented in large-scale automatic segmentation tools, will be essential in order to accurately capture the sub-field nature of subcortical nuclei. Promising new tools like \textit{NextBrain} \citep{casamitjana2024next} could help implementing updated label definitions in the future for automatic UHF segmentation techniques like \go{}.

\subsection{Limitations}

It is important to acknowledge certain limitations of our study, such as the limited availability of reference techniques to compare \go{} to. As discussed in section \ref{disc:ceb}, it would have been preferable if \ceb{} would have offered the same out-of-the-box implementation like \fsv{} or \sys{} (albeit the extra external up-sampling step that had to be added by the authors in order to enable comparisons for \sys{}). Additionally, it is worth repeating that \ceb{} produces only 6 labels whereas \go{} produces 35 (following \fss{}/\fsv{} label convention). This allows for a considerably larger number of regions to use for quantitative analyses with \go{}, which is especially of interest at UHF-MRI. Ultimately, this lack of reference segmentation techniques at UHF-MRI further manifests the need for novel techniques to be developed.

A common issue faced by all novel segmentation techniques is the sparsity of real ground truth segmentations to use for testing. In this work, manual segmentations were only available for the UltraCortex dataset and two labels only. For other quantitative analyses, either \fsv{} segments computed on \tow{} images or \sys{} were used as a "silver standard" or, for \ceb{}, the iGT was used. To the best of our knowledge, no dataset available online offers 3D sub-millimeter manual segmentations for several subcortical labels at UHF-MRI. Moreover, producing our own manual segmentations would have been extremely time-consuming and required expertise outside the scope of this work. In addition, manual segmentations are prone to inter- and intra-expert variability \citep{deeley2011comparison}. 

Despite \go{} being able to segment any contrast and resolution tested, input images still need to be skull-stripped unlike similar techniques (\fsv{}, \sys{} or \ceb{}). The reasons behind this requirement are that, first, some training data were already skull-stripped when accessed, and second, segmenting extra-cerebral labels as in \cite{billot2023synthseg} was not easily obtainable for UHF images due to signal inhomogeneities outside the brain. Extra-cerebral labels are required in order to generate synthetic contrasts for the whole head and such tools are not readily available for UHF images. On the other hand, considering that skull-stripping is a quite common step for neuroimaging pipelines, and that it has been extensively developed and improved recently with the arrival of DL-based techniques, we strongly believe that it should not limit the usability of \go{} in practice. Indeed, a wide variety of robust and extensively tested options are freely and easily available online like BET, HD-BET, SynthStrip, ROBEX, ANTsPyNet, MONSTR, etc. \citep{smith2000bet, isensee2019automated, hoopes2022synthstrip, iglesias2011robust, roy2017robust, tustison2021antsx}. Nonetheless, possible errors in skull-stripping can impact the quality of the segmentation results and we recommend users to assess the skull-stripping on their images before using \go{} and use a consistent procedure for a given image type.

A potential drawback of \go{}, designed for UHF-MRI, is the fact that cortex parcellation is not performed. This can be a limitation for researchers using functional MRI (fMRI) at UHF where its advantages, compared to 3T, have been shown \citep{beisteiner2011clinical}. In addition, as previously mentioned, all training and most of the test data in this study consisted of MR images of young healthy subjects. While the volumetry results on PDP versus HC indicate that \go{} is comparable to \fsv{}, they also pointed to some potential problems of \go{} related to enlarged lateral ventricles. A broad and systematic evaluation of the performance of \go{} in the presence of deviating anatomies and various pathologies was outside the scope of the current study, but should be done in a separate follow-up study. Such a study should also consider a retraining of \go{} with added data from patient studies in the training corpus. This could, for example, include open-access databases with neurological disorders like the OASIS or ADNI databases \citep{marcus2007open,jack2008alzheimer}.  Originally excluded due to their lower resolutions (i.e., 1 mm$^3$), these datasets could offer anatomical variations, like enlarged ventricles, that can be impossible to synthesize with the generative model and, thus, improve the robustness of \go{} to a wider range of brain anatomies. Ultimately, addressing the limitations related to the cortex parcellation, lack of anatomical diversity in the training data and thorough testing of \go{} on clinical cohorts with pathologies represents the main focus of future work.

\section{Conclusions}\label{conc}

In summary, we propose \go{}, a novel DL-based segmentation technique capable of segmenting MR images acquired with various contrasts, resolutions and even field strengths. \go{} was able to segment all six resolutions and seven contrasts tested in this work. The usage of synthetic images for training enabled the segmentation of images acquired at 3T, 7T and 9.4T. At 3T, when compared to \fsv{}, \go{} gave an average DSC of 0.89 for both \tow{} and \ttw{} images, demonstrating great performance at lower field strengths and superiority over \sys{}, although developed for UHF applications. At 7T, \go{} was able to segment five different contrasts and showed similar performance to \ceb{} while being substantially more generalizable and practical for the UHF-MRI context. At 9.4T, \go{} demonstrated high agreement with manual segmentations with an average DSC of 0.93 over 12 subjects versus 0.89 for \sys{}. Despite \sys{} exhibiting decent performance at UHF with high inhomogeneity resistance, \sys{} lacked the necessary granularity required at UHF in its output segmentations, likely due to the low training resolution. Ultimately, by being trained on synthetic images randomly generated from only sub-millimeter label maps, \go{} was able to develop contrast- and resolution-agnostic capabilities adapted to the UHF-MRI reality with, in addition, a significant resistance to noise and signal inhomogeneities, which have been a major challenge for automatic segmentation at UHF-MRI until now. For this initial version of \go{}, the training and testing were predominantly conducted using data from healthy subjects. While this will be addressed in its next iteration, it is important to consider this factor when applying \go{} to patient cohorts.

\section*{Data and Code Availability}

The source code for \go{} is available on GitHub at \href{https://github.com/mafortin/GOUHFI}{https://github.com/mafortin/GOUHFI}.

All MRI datasets used in this article are open source repositories freely available through the links provided in footnotes of section \ref{meth} except for the \sca{} and STRAT-PARK datasets. The latter are not publicly available due to data protection regulations.

\section*{Author Contributions}

\textbf{MAF}: Conceptualization, Data Curation, Methodology, Formal analysis, Investigation, Software, Validation, Writing-original draft, Writing-review \& editing, Visualization. \textbf{ALK}: Writing-review \& editing, Data Curation, Validation, Formal analysis. \textbf{MSL}: Conceptualization, Writing-review \& editing. \textbf{LL}: Data acquisition, Writing-review \& editing. \textbf{RS}: Data acquisition, Writing-review \& editing. \textbf{PEG}: Supervision, Resources, Project administration, Writing-review \& editing, Funding acquisition.

\section*{Funding}

This project and \sca{} data acquisition were supported through the following funding organizations under the aegis of JPND: Belgium, The Fund for Scientific Research (F.R.S.-FNRS; funding code PINT-MULTI/BEJR.8006.20); Germany, Federal Ministry of Education and Research (BMBF; funding codes 01ED2109A/B); and Norway, The Research Council of Norway (RCN; funding code 322980).

\section*{Declaration of Competing Interests}

The authors do not declare any competing interests.

\section*{Acknowledgments}

First, we would like to thank the SCAIFIELD consortium (Principal Investigator: Tony Stöcker) and the STRAT-PARK study (Study directors: Charalampos Tzoulis \& Mandar Jog) for supporting the data acquisition and funding related to each dataset used in this article. \\
We would also like to thank all the publicly available datasets used in this work. First, data provided as part as the Human Connectome Project, WU-Minn Consortium (Principal Investigators: David Van Essen and Kamil Ugurbil; 1U54MH091657) funded by the 16 NIH Institutes and Centers that support the NIH Blueprint for Neuroscience Research; and by the McDonnell Center for Systems Neuroscience at Washington University were used in this publication. Finally, we would also like to thank individually all these following projects/initiatives for sharing their dataset and making this article possible: (1) the UltraCortex: Submillimeter Ultra-High Field 9.4 T1 Brain MR Image Collection and Manual Cortical Segmentations dataset (DOI:\href{https://doi.org/10.18112/openneuro.ds005216.v1.1.0}{https://doi.org/10.18112/openneuro.ds005216.v1.1.0}), (2) Autism Brain Imaging Data Exchange II (ABIDE-II) initiative (DOI: \href{https://dx.doi.org/10.21227/y3v9-b041}{https://dx.doi.org/10.21227/y3v9-b041}), (3) Open Science CBS Neuroimaging Repository (\href{https://www.nitrc.org/frs/?group_id=606}{https://www.nitrc.org/frs/?group\_id=606}), (4) CEREBRUM-7T: Fast and Fully-volumetric Brain Segmentation of 7 Tesla MR Volumes (DOI:\href{https://doi.org/10.25493/RF12-09N}{https://doi.org/10.25493/RF12-09N}) and (5) Human Brain Atlas (\href{https://osf.io/ckh5t/}{https://osf.io/ckh5t/}). 


\printbibliography

\section{Appendix}
\subsection{List of the structures and corresponding label values segmented by \go{}}
\label{appA}

\begin{table}[h!]
\centering
\begin{tabular}{l r}
\hline
\footnotesize 
\textbf{Segmented structure} & \textbf{Label index} \\
\hline
Cerebral white matter (lh) & 1 \\
Cerebral cortex (lh) & 2 \\
Lateral Ventricle (lh) & 3 \\
Inferior Lateral Ventricle (lh) & 4 \\
Cerebellar White Matter (lh) & 5 \\
Cerebellar Cortex (lh) & 6 \\
Thalamus (lh) & 7 \\
Caudate (lh) & 8 \\
Putamen (lh) & 9 \\
Pallidum (lh) & 10 \\
3rd-Ventricle & 11 \\
4th-Ventricle & 12 \\
Brain Stem & 13 \\
Hippocampus (lh) & 14 \\
Amygdala (lh) & 15 \\
CSF & 16 \\
Accumbens (lh) & 17 \\
Ventral DC (lh) & 18 \\
Choroid Plexus (lh) & 19 \\
Cerebral white matter (rh) & 20 \\
Cerebral cortex (rh) & 21 \\
Lateral Ventricle (rh) & 22 \\
Inferior Lateral Ventricle (rh) & 23 \\
Cerebellar White Matter (rh) & 24 \\
Cerebellar Cortex (rh) & 25 \\
Thalamus (rh) & 26 \\
Caudate (rh) & 27 \\
Putamen (rh) & 28 \\
Pallidum (rh) & 29 \\
Hippocampus (rh) & 30 \\
Amygdala (rh) & 31 \\
Accumbens (rh) & 32 \\
Ventral DC (rh) & 33 \\
Choroid Plexus (rh) & 34 \\
WM-hypointensities & 35 \\
Extra-Cerebral & 36$^*$ \\
\hline
\end{tabular}
\caption{Label name and index of all structures segmented by \go{}. Lh and rh stands for Left- and Right-Hemisphere respectively. \textbf{$^*$}: While required to generate the synthetic images used for training, the Extra-Cerebral label is not segmented by \go{}.}
\label{table:labels}
\end{table}

\clearpage
\subsection{Parameters of the generative model for synthetic image creation}
\label{appB}

\begin{table}[h]
\centering
\begin{tabular}{l r}
\hline
\textbf{Parameter} & \textbf{Value} \\
\hline
a$_{rot}$ & -20 \\
b$_{rot}$ & 20 \\
a$_{sc}$ & 0.8 \\
b$_{sc}$ & 1.2 \\
a$_{sh}$ & -0.015 \\
b$_{sh}$ & 0.015 \\
a$_{tr}$ & -30 \\
b$_{tr}$ & 30 \\
b$_{nonlin}$ & 4.0 \\
a$_\mu$ & 0 \\
b$_\mu$ & 255 \\
a$_\sigma$ & 0 \\
b$_\sigma$ & 35 \\
b$_B$ & 0.9 \\
$\sigma^{2}_\gamma$ & 0.4 \\
r$_{HR}$ & None \\
b$_{res}$ & None \\
a$_\alpha$ & None \\
b$_\alpha$ & None \\
\hline
\end{tabular}
\caption{Values of the parameters used in this study for the generative model. Intensity parameters assume inputs in the [0, 255] interval, rotations are
expressed in degrees with spatial measures in millimeters. More details about these parameters are provided in \cite{billot2023synthseg}.}
\label{table:hyperparameters}
\end{table}

\clearpage
\subsection{\fss{} vs \fsv{} label maps for \sevi{} 3T \mpr{} images}
\label{appC}

\begin{figure*}[hbt!]
    \centering
    \includegraphics[width=\textwidth]{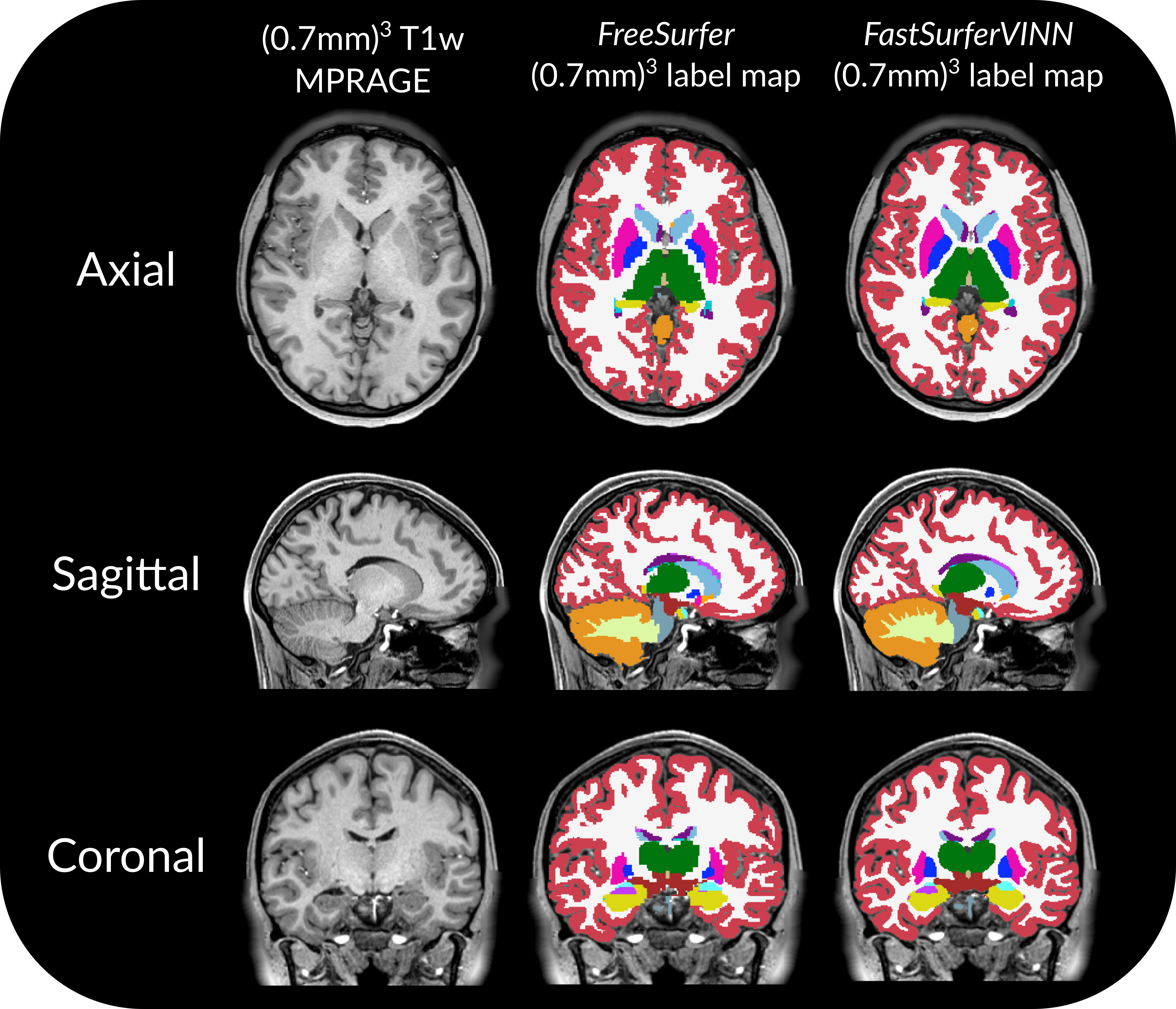}
    \caption{Comparison in the three anatomical planes of the output label maps produced by \fss{} (middle column) and \fsv{} (right column) for the same 3T \sevi{} \tow{} \mpr{}. Despite both label maps being at the same resolution of \sevi{}, a difference in delineation quality is easily observable between \fss{} and \fsv{}, with the latter producing more refined delineations of anatomical structures. Therefore, \fsv{} was preferred for the training label maps for \go{}.}
    \label{fig:appC}
\end{figure*} 

\end{document}